\documentclass[12pt]{article}
\usepackage{graphics, color}
\usepackage{graphicx}
\usepackage{amssymb}
\usepackage{latexsym}
\usepackage{amsthm}
\usepackage{amsmath}
\usepackage{slashed}
\usepackage{hyperref}
\usepackage{enumerate}
\numberwithin{equation}{section}
\pdfoutput=1

\def\gsim{\, \rlap{$>$}{\lower 1.1ex\hbox{$\sim$}}\,}
\def\lsim{\, \rlap{$<$}{\lower 1.1ex\hbox{$\sim$}}\,}

\newcommand{\comment}[1]{}
\newcommand{\expect}[1]{\left\langle #1 \right\rangle}
\newcommand{\mbf}[1]{\mathbf #1}
\newcommand{\mal}{\mathcal}
\def\e{\mathrm{e}}

\def\d{\partial}

\def\x{\mbf x}
\def \k {\mbf k}
\def\q{\mbf q}
\def\p {\mbf p}
\def \O{\mal O}

\newcommand{\be}{\begin{equation}}
\newcommand{\ee}{\end{equation}}
\newcommand{\bea}{\begin{eqnarray}}
\newcommand{\eea}{\end{eqnarray}}

\def\mpl{M_{\rm Pl}}

\begin{document}

\begin{titlepage}

\setcounter{page}{1} \baselineskip=15.5pt \thispagestyle{empty}

\begin{flushright}
{\footnotesize{SU/ITP-16/12, UTTG-09-16}} \date{}
\end{flushright}


\vspace{0.5cm}
\begin{center}
{\fontsize{21}{34}\selectfont  \sc
Productive Interactions: \\[.3cm] heavy particles and non-Gaussianity
\vspace{4mm}
}
\end{center}

\begin{center}
{\fontsize{13}{30}\selectfont Raphael Flauger,$^1$ Mehrdad Mirbabayi,$^2$ Leonardo Senatore,$^{3,4,5}$ and Eva Silverstein,$^{3,4,5}$ }
\end{center}

\begin{center}
\vskip .6cm

\textsl{$^1$ Department of Physics, The University of Texas at Austin, Austin, TX, 78712, USA}

\vskip 7pt
\textsl{$^2$ Institute for Advanced Study, Princeton, NJ 08540, USA}

\vskip 7pt
\textsl{$^3$ Stanford Institute for Theoretical Physics, Stanford University, Stanford, CA 94305, USA}

\vskip 7pt
\textsl{$^4$ SLAC National Accelerator Laboratory, 2575 Sand Hill Rd., Menlo Park, CA 94025, USA}

\vskip 7pt
\textsl{$^5$ Kavli Institute for Particle Astrophysics and Cosmology, Stanford, CA 94305, USA}


\end{center}

\vspace{.5cm}
{ \noindent \textbf{Abstract} \\[0.1cm]
\hrule \vspace{0.3cm}
\noindent We analyze the shape and amplitude of oscillatory features in the primordial power spectrum and non-Gaussianity induced by periodic production of heavy degrees of freedom coupled to the inflaton $\phi$.  We find that non-adiabatic production of particles can contribute effects which are detectable or constrainable using cosmological data even if their time-dependent masses are always heavier than the scale $\dot \phi^{1/2}$, much larger than the Hubble scale.  This provides a new role for UV completion,  consistent with the criteria from effective field theory for when heavy fields cannot be integrated out.    This analysis is motivated in part by the structure of axion monodromy, and leads to an additional oscillatory signature in a subset of its parameter space.  At the level of a quantum field theory model that we analyze in detail, the effect arises consistently with radiative stability for an interesting window of couplings up to of order $\lesssim 1$.  The amplitude of the bispectrum and higher-point functions can be larger than that for Resonant Non-Gaussianity, and its signal/noise may be comparable to that of the corresponding oscillations in the power spectrum (and even somewhat larger within a controlled regime of parameters).   Its shape is distinct from previously analyzed templates, but was partly motivated by the oscillatory equilateral searches performed recently by the {\it Planck} collaboration.    We also make some general comments about the challenges involved in making a systematic study of primordial non-Gaussianity.  }
\vspace{0.3cm}
 \hrule

\end{titlepage}

\tableofcontents
\vspace{0.3cm}
\newpage

\baselineskip = 16pt

\section{Introduction}

The observation of the primordial seeds for structure provides a fertile testing ground for theories of the dynamics that generates them.  
The theoretical and observational study of the primordial power spectrum and non-Gaussianity is a mature field, with substantial progress recently due to {\it Planck} \cite{Planck15NG}\ and future possibilities in large-scale structure.  But even in the CMB this study is not complete; in fact it is not known if there is a systematic way to complete it. 

In principle, there is an infinite space of possibilities, in practice the only useful searches involve $N$-spectra which depend on a limited number of parameters.  Such templates can be derived from sufficiently well-defined theories of the primordial perturbations. Distinct analyses are required in order to test shapes which do not overlap strongly, in the precise sense developed in~\cite{shapes}.     Moreover, a physical mechanism which generates large non-Gaussianity may generically also affect the power spectrum, so such searches are only well-motivated if the signal/noise in the higher-point functions is competitive with the leading corrections to the power spectrum.  

In this work, we will present a new class of shapes motivated by a very basic theoretical possibility:
non-derivative couplings of the inflaton to additional heavy fields. In the presence of a discrete shift symmetry such couplings can be significant without spoiling inflation
and, as we will see, can lead to non-adiabatic production of very heavy fields, sourcing detectable corrections to the scalar perturbations -- including non-Gaussianity -- in an interesting range of parameters.  

Such couplings occur in axion monodromy inflation \cite{monodromy, ignoble, recentmonodromy, unwinding} at the single-light field level, and were investigated in \cite{SSZ,MSSZ} as a source of tensor emission during inflation (see e.g. \cite{CookSorbo, othersecondary} for another secondary source of tensor modes). The scalar contribution was a limiting factor on this effect.   Here we analyze this contribution in detail, focusing on the regime where the scalar emission is subdominant to the nearly Gaussian vacuum fluctuations, but can be detectable in the power spectrum and non-Gaussianity.   In the regime we consider, the particle production does not backreact on the inflationary dynamics, in contrast to \cite{trapped}, where the perturbations were studied in a continuum approximation in the strongly back-reacting regime.

In broader terms, we will show how microscopic subhorizon physics during inflation can be relevant for the superhorizon predictions when the inflaton is coupled to additional fields, even very heavy ones. 

\begin{figure}[htbp]
\begin{center}
\includegraphics[width=12cm]{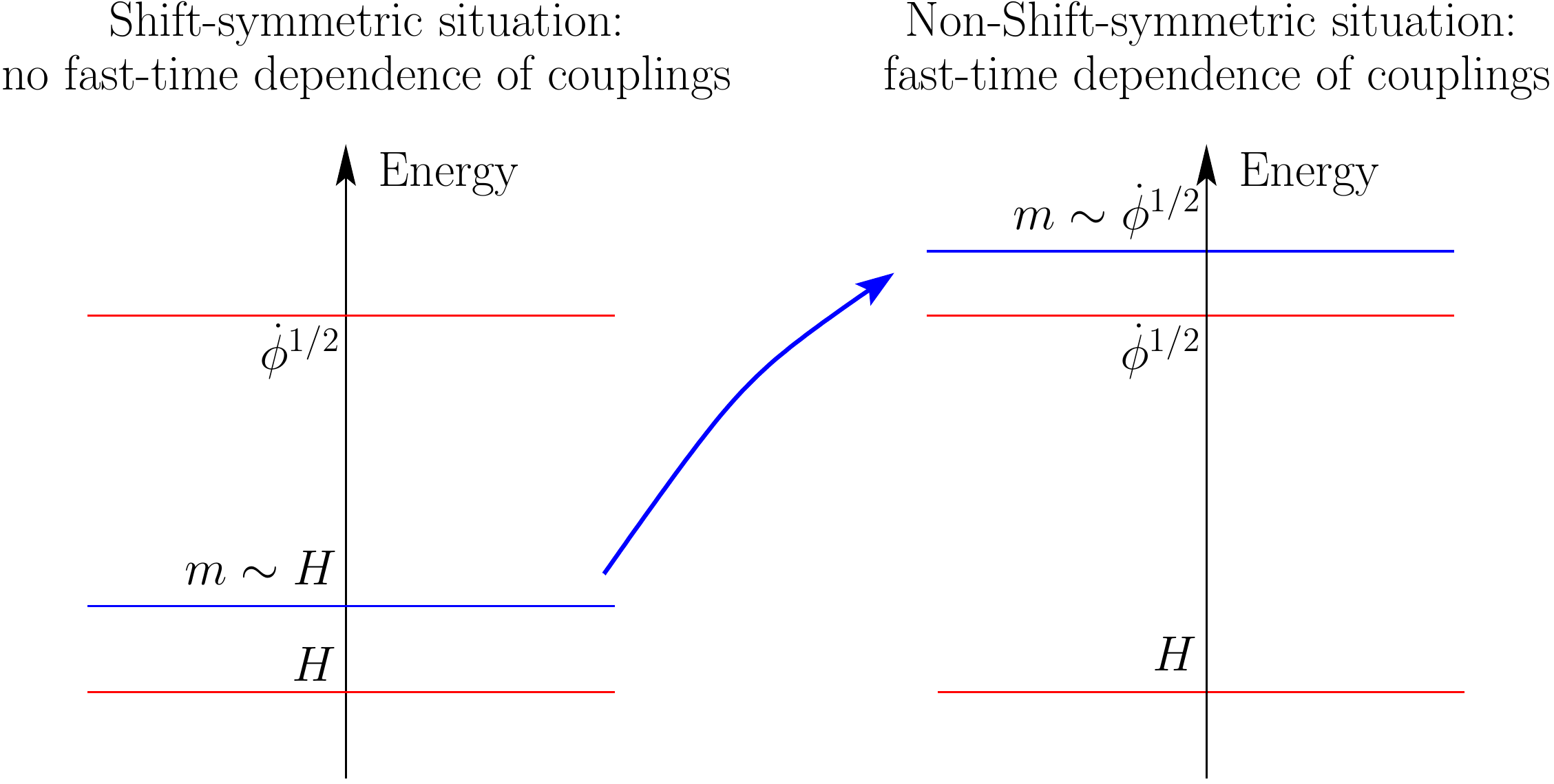}
\end{center}
\caption{Pictorial representation of our findings: in an inflationary theory with an approximate continuous shift symmetry for the inflaton, only particles that are not much heavier than the Hubble scale $H$ are relevant for the dynamics of the fluctuations.  However, as we will see, if the continuous shift symmetry is broken, e.g. to a discrete shift symmetry, heavier particles can become relevant as depicted on the right. In the scenarios studied in this work, the new scale is set by $\dot\phi$.  The basic estimate $\exp(-\pi m^2/\dot\phi)\sim 1/\sqrt{N_{\rm modes}}$ suggests observational sensitivity to these massive particles, which we confirm in a detailed analysis.}
\label{fig:mass_scales}
\end{figure}

We will exhibit two novel effects:

$\bullet$  Even fields that are never lighter than the scale $\dot\phi^{1/2}$ in slow-roll inflation\footnote{more generally, the scale $\dot m^{1/2}$ of the time dependence in the heavy particle masses.} can make a measurable difference to the primordial $N$-spectra.  From the perspective of the  effective field theory of inflaton fluctuations \cite{EFT, resonantNG, resonantEFT}, the dynamical scales that are visible are the expansion rate $H$ and the Fourier frequency of the time dependent mass and couplings.  We will see that fields comparable to these scales can modify the primordial $N$-spectra in such a way that is not captured by any effective single field description. There has been much interesting previous work on the potentially observable effects of thermally produced massive particles in the expanding background, such as \cite{massiveH}, and the effects of heavier fields on the underlying inflationary dynamics  (e.g. \cite{flattening}). The latter explains inflationary plateaus,\footnote{and clarifies the continued viability of string-theoretic inflation mechanisms \cite{stringreviews}, given the presence of heavy fields which adjust to suppress the inflationary potential energy.} but does not provide a distinctive signature. The former leads to effects suppressed exponentially in $m/H$ for mass $m$. As we will see, the present mechanism will generate a less suppressed effect of heavy fields, with the number density of the of heavy particles and hence the amplitude governed by the exponential factor 
\be\label{expform}
\exp(-\pi\tilde\mu^2/g\dot\phi)
\ee
where $\tilde\mu$ is the lightest value of the time dependent mass in question, and $g$ a coupling which can be order 1. Since $\dot\phi\sim 58^2 H^2$ this has much greater amplitude for a given mass than the effects arising purely from vacuum fluctuations.  This result is pictorially represented in Fig.~\ref{fig:mass_scales}.

$\bullet$  For most oscillation frequencies $\omega$ of interest, the amplitude of the resulting oscillatory non-Gaussianity can be parametrically larger than that of previously studied resonant non-Gaussianity \cite{resonantNG}\cite{resonantEFT}, derived from a slow roll potential with a small sinusoidal term. In fact, for rare events (when the factor \eqref{expform} becomes small), in the regime where the coupling $g$ is not too small, the present mechanism produces highly non-Gaussian perturbations with signal/noise easily competitive with that of the oscillatory features in the power spectrum.  This provides theoretical motivation for a joint analysis of such templates in the power spectrum and bispectrum, analogous to \cite{jointpowerbi}. The contrast between the present effect and resonant non-Gaussianity is expressed in the simple formulas (\ref{three2tworesonant}) and (\ref{three2two}) below.    In fact, we find that the signal/noise in the primordial $N$ point correlators can grow with $N$ for a range of $N$; in this regime it would be interesting to determine the optimal search strategy.

In~\S\ref{setup} we will review the setup and radiative stability of the model. In~\S\ref{power}, we will calculate the correlation functions of scalar perturbations that result from particle production. We discuss the constraints on the model parameters enforced to stay within the regime of validity of our approximations in~\S\ref{sec:parameterwindows}. The phenomenology of the model and the templates for analysis are laid out in \S\ref{templates}.  Finally, in \S\ref{general} we will make some general remarks about how our results compare to previous mechanisms generating significant non-Gaussianity, and comment on the  interplay between the time-dependent couplings in the EFT of perturbations, non-adiabatic effects, and data searches before concluding in \S\ref{summary}.


\section{Setup and vacuum loop corrections}\label{setup}

We will be interested in the coupling of one or more heavy fields $\chi_I$ to the inflaton $\phi$, leading to a field-dependent mass
\be\label{phichi}
V(\chi_I, \phi) \simeq \sum_I \frac{1}{2}m_{\chi_I}^2(\phi) \chi_I^2 + V_0(\phi)\,,
\ee
which implies a time-dependent mass for $\chi_I$ as $\phi$ rolls during inflation. If this time-dependence is sufficiently rapid, it leads to non-adiabatic production of $\chi_I$ particles. The produced particles then source inflaton fluctuations as their mass changes in time. We will find that even $\chi$ fields which are never lighter than $\dot\phi^{1/2}$ can contribute measurably to perturbations in some regime of parameters and hence cannot be integrated out. 

There are many ways such couplings can appear -- in general for each such field there could be an arbitrary mass function.  We will develop this in some generality, but ultimately focus on couplings respecting an approximate discrete shift symmetry, weakly broken by the slow roll potential, as motivated by axions.  Although that narrows down the possibilities, postulating this symmetry does not suffice to determine the observables, as we will see explicitly, and more theoretical input is required.   
A mass function that is disordered as discussed recently in \cite{disorder} is another interesting limit.  
          
The structure of axion monodromy in string theory motivates the discrete shift symmetry, and entails further sectors of fields and couplings.  In that theory, there
are two types of heavy $\chi$ sectors with masses $m_\chi(\phi)$ that arise from the same basic structure and are specific enough to derive concrete oscillatory $N$-spectrum shapes and amplitudes.    

\smallskip

\noindent{\bf (a) particle sectors with monodromy structure}

There is a part of the spectrum which undergoes monodromy in analogy to the potential energy, with a different sector reaching a minimal mass or tension each time the field traverses an underlying period in the axion field space.  If these degrees of freedom are particles (as opposed to strings), we have
\be\label{2amasses}
m_{\chi_n,(a)}^{2}=\mu_a^2+ \hat{\mu}_a^2(a(\phi)-2\pi n)^2 \simeq \mu_a^2+g_a^2(\phi-2\pi n f)^2
\ee
where $a(\phi)$ is the periodic axion field as a function of the canonically normalized inflaton field $\phi$.  In the absence of drift \cite{drift}, $a(\phi)=\phi/f$ with an axion decay constant $f$, leading to the last expression in (\ref{2amasses}) with $g_a=\hat{\mu}_a/f$.  

As just mentioned, depending on the microphysical details the produced sources may be extended strings rather than particles.  Non-adiabatic production of strings has some very interesting subtleties and potentially distinguishing features \cite{SSZ}\cite{backdraft}.  To be specific, we will focus on the particle production case in this paper.


\noindent{\bf (b) sinusoidally modulated masses}

Additional fields coupled to the inflaton generically have masses modulated by the periodic term in the potential. For example, as the axion traverses its underlying period, massive fields, such as moduli or Kaluza-Klein modes, will undergo periodic modulation of their masses. This motivates a $\chi$ mass-squared of the form
\be\label{bmasssquaredgf}
m_\chi^2=\mu^2+2 g^2 f^2 \cos\frac{\phi}{f}
\ee
and requires
\be\label{gfmuineq}
gf<\frac{\mu}{\sqrt{2}}.
\ee
Near a point of minimal mass, where the argument of the cosine is $\pi(2n+1)$, this behaves as
\be\label{minmassb}
m_{\chi,(b)}^2=\mu^2-2 g^2 f^2+{g^2}(\phi-\phi_n)^2+\dots \equiv \mu_b^2+{g^2}(\phi-\phi_n)^2+\dots
\ee
In the regime we will be interested in, the form (\ref{minmassb}) governs the physics near a particle production point as we will explain shortly. Although this regime is similar to case (a), the sinusoidal behavior of the resulting source of $\delta\phi$ perturbations will lead to distinct results for the scalar perturbations in case (b). In both cases the angular frequency of production events is given by 
\be\label{omega}
\omega = \frac{\dot\phi}{f}.
\ee

\subsection{EFT perspective and dependence on high energy scales}\label{EFT} 

Before proceeding to analyze this mechanism in detail, let us address how the need to incorporate the heavy fields arises in the effecitve field theory context \cite{EFT}.   One of the lessons of the present work will be that the precision of current data can require inclusion of very heavy fields which one cannot trivially integrate out.
If the EFT Lagrangian has some time dependent couplings $M_{i}^4(t)$ and $H(t)$ in the action  (see equation (\ref{EFTLagrangian}) below), and if their Fourier transforms have support at frequency of order $\hat\omega$,  any additional particles that exist with mass $m$ of order $\hat\omega$ will be produced. 

Before moving to the present work, let us first note that this fact is already familiar in standard inflationary scenarios where the time dependence induced by the Hubble scale in the metric induces fluctuations in particles of mass $m\lesssim {\cal O}(H)$. As particles get parametrically heavier than $H$, one might naively imagine reducing to a single-field EFT by integrating out such particles. But the resulting EFT would only include all terms in an analytic expansion in $H^2/m^2$; non-analytic effects in this parameter, for example those which scale as $e^{-m/H}$, would be missed. 
Depending on the precision of the experiment, these non-perturbatively small effects may not be negligible.

Exactly the same considerations apply to all time-dependent  couplings in the Lagrangian for the EFT of inflationary perturbations (\ref{EFTLagrangian}). If the EFT Lagrangian contains functions of time with support at frequencies $\hat\omega$ of order the mass $m$, a naive single-clock theory would miss effects that scale as $e^{-m/\hat\omega}$.  

It is very possible for the effective theory of perturbations to have functions of time that have support at frequencies of order $\hat\omega\sim \dot\phi$: even in slow-roll inflation, time-translations are broken at the scale $\dot\phi^2\sim \dot H\mpl^2\gg H^4$. Let us apply this, for example, to slow roll inflation in the presence of a particle whose mass depends on the inflaton as $m(\phi(t))\sim \sqrt{\mu^2+\dot\phi^2 t^2}$. If we were  to integrate out this particle, we would get contributions to the low energy theory proportional to $M_i^4(t),H(t)\propto 1/m(\phi(t))^2\sim \frac{1}{\mu^2+\dot\phi^2 t^2}$ whose Fourier transform scales as $e^{-\omega/\sqrt{\dot\phi}}$ (taking $\mu\sim\sqrt{\dot\phi}$).  This has support up to frequencies of order $\dot\phi^{1/2}$.  


As we will see in detail in this paper, this implies that particles with mass $\sim \dot\phi^{1/2}$, and even slightly larger than this, can be produced at a detectable level. In fact, a cosmological experiment that measures $N_{\rm modes}$ cosmological modes, has a relative precision of about $N_{\rm modes}^{-1/2}$; for an experiment such as {\it Planck}, this is about $10^{-3}$. This allows for the exponent of the exponential suppression to be large (with the details depending on additional power law factors as we will see in this work). 

Notice that there are no surprises from the EFT point of view: in all EFT's, the number of relevant degrees of freedom should be declared {\it a priori}. If the time-dependent couplings have a Fourier transform with support at frequencies of order $\hat\omega$, insisting on a single-clock description of inflation amounts to assuming that no additional particles are present with mass lighter than $\hat\omega$ (and even a bit larger than $\hat\omega$ because we can afford for some exponential suppression). This depends on the full model of inflation.  As such, this provides a new role for UV completion that goes beyond simply deriving the light field spectrum from it and controlling Planck-suppressed corrections to the inflationary dynamics.  

One might wonder how effects that are exponentially suppressed as $e^{-m^2/\dot\phi}$ can dominate over the ones that come from integrating out these particles. As we will see in our detailed analysis below, this can be understood as follows.   Integrating out heavy particles in slow roll inflation, for example, will induce operators such as $(\d\phi)^4/m^4\sim \dot\phi_0 (\d\delta\phi)^3/m^4$. This operator induces a non-Gaussianity with signal to noise ratio $S/N\sim f_{\rm NL}\zeta\sim (\dot\phi_0^2/m^4)\zeta$ (as can be read off from the ratio of the three point interaction Lagrangian to the kinetic term, ${\cal L}_3/{\cal L}_2$) \cite{Creminelli}. This is $\lesssim \zeta\sim 10^{-4}$ if $m\gtrsim \dot \phi$.  The non-Gaussianity from particle production works differently.  As we will review in detail below (see e.g. \cite{KLS, trapping}), it induces a number of particles in an Hubble patch, $N_H$, (which is the relevant quantity) that scales as $N_H\sim (\dot\phi^{1/2}/H)^3 e^{-m^2/\dot\phi}$, where we have just kept the most important parametric dependence. The non-Gaussianity of this distribution is controlled by $N_H^{-1/2}$, which can be order one for $N_H\sim 1$. This non-Gaussianity will then be transferred to the inflaton through the relevant couplings. The prefactor $ (\dot\phi^{1/2}/H)^3\sim 10^6$ is a very large number, allowing for some exponential suppression to be present, while preserving the dominance of the effect.

In short, the observable fluctuations during inflation can be sensitive to scales that are as much as a few orders of magnitude higher than $H$. If $H$ happens to be sufficiently high, this corresponds to scales close to the GUT scale or so, which makes the possibility even more interesting.

\subsection{Vacuum loop corrections}\label{loopsbackground}

The coupling of the $\chi$ fields to the inflaton $\phi$ generates two types of corrections to the dynamics 
of $\phi$, which can roughly be characterized as those coming from $\chi$ loops (vacuum fluctuations) and those coming from $\chi$ production; of course in general there is a combination of the two.  The latter effects are the main subject of this paper.  The former must be taken into account as well.  Their size depends on microscopic details such as the level of broken supersymmetry in the $\chi-\phi$ sector.    The contribution from $\chi$ vacuum fluctuations generates various periodic terms in the effective action.  This may be the leading such contribution, or it may be subdominant, depending on parameters.

In general, before computing the effects of particle production, we should ensure that the system we are considering is controlled against radiative corrections. The condition for radiative stability is important to our assessment of the non-Gaussianity:  it restricts the strength of the coupling constant in the theory, and hence leads to some constraint on the strength of nonlinear interactions visible in the perturbations. In this regard, it is worth emphasizing that with microscopic supersymmetry, the contributions of bosons and fermions can (partially) cancel each other in the corrections to the effective action, whereas they arise additively in the non-adiabatic effects that we are concerned with in this paper. Therefore, we analyze radiative corrections in appendix \ref{app:radiative} assuming some degree of supersymmetry. We will find that for the corrections to the slow-roll potential to be subdominant we need
\be\label{g}
g\ll 4\pi,
\ee
and for the higher derivative corrections not to induce large non-Gaussianity (as in DBI inflation \cite{DBI})
\be\label{veff}
v^2=\frac{g^2}{2}N_3\left(\frac{g\dot\phi}{\pi\mu^2} \right)^2 \ll 1\qquad\text{with}\qquad N_3 = \mu /(2\pi gf)\,.
\ee
Another way to see the existence of a regime of sub-dominance of the non-Gaussianity induced by power law corrections to effective action was described above in \S\ref{EFT}.  

\section{Power spectrum and non-Gaussianity from particle production}\label{power}

In this section, we will derive the shape and amplitude of the contribution to the power spectrum and bispectrum (as well as higher-point correlators) from repeated particle production events. We will first give a detailed description of the non-adiabatic production and evolution of the heavy particles (sources) in \S\ref{sec:prod}. In \S\ref{sec:sourced}, we will derive the spectrum of classical scalar emission by these sources.\footnote{This shape was partly computed in appendix B of \cite{MSSZ}, where it was a limiting factor for alternative sources of tensor modes \cite{SSZ}.} In \S\ref{higherbetasquared}, we include the effect of interaction with vacuum scalar fluctuations. Finally, in \S\ref{sec:interference} we consider interference terms between two branches of the wave-function: one without particle production and one in which a pair of $\chi$ particles are produced and subsequently annihilate into $\delta\phi$ quanta. 

\subsection{Source dynamics}\label{sec:prod}

We will work in a regime where the timescale $t_{pr}$ associated with each production event is shorter than half the time period $\sim 2\pi/\omega$ separating the events.  Each production event is then well modeled by a time-dependent mass of the form 
\be\label{localmodel}
m_\chi^2(t)|_{|t-t_n|\lesssim t_{pr}}=\tilde\mu^2 + g^2(\phi-\phi_n)^2 \simeq \tilde\mu^2+g^2\dot\phi^2 (t-t_n)^2
\ee
where $\tilde\mu=\mu_a$ or $\mu_b$ in our two specific cases described above.  Here $n$ labels the event, and in the regime $\tilde\mu^2 > g\dot\phi$ the timescale on which the production occurs is
\be\label{tpr}
t_{pr}\sim \frac{\sqrt{2}\tilde\mu}{g\dot\phi}\,.
\ee
This follows from maximizing $\dot m_\chi/m_\chi^2$ as a function of $|t-t_n|$, giving $t-t_n=\pm \tilde\mu/(\sqrt{2}g\dot\phi)$.  Particle production including cosmological applications has been discussed extensively in the literature; see for example \cite{KLS}\cite{otherproduction}.     

Before proceeding further, let us check that in case (b) we can indeed obtain an inequality allowing us to model the production event using (\ref{localmodel}).  This requires 
\be\label{tprineq}
t_{pr}< \frac{\pi}{\omega} \Rightarrow gf>\frac{\sqrt{2} \tilde\mu}{\pi}\,,
\ee
where we used equations (\ref{omega}) and (\ref{tpr}).  This is consistent with the basic requirement  
(\ref{gfmuineq}) above in case (b), since there is a window
\be\label{gfwindow}
\mu_b^2 \frac{2}{\pi^2} = (\mu^2-2 g^2 f^2)  \frac{2}{\pi^2}< g^2 f^2 < \frac{\mu^2}{2}\,.
\ee
To fix our conventions and notation, the inflationary metric is approximately de Sitter
\be\label{dS}
ds^2=-dt^2+a^2(t) d\x^2 = a^2(\eta)(-d\eta^2+d\x^2)\,,
\ee
with conformal time coordinate $\eta=-1/aH$.  
Let us denote the comoving momenta by $\k$ and physical momenta by $\p$, which are given at the time of the $n$-th production event by
\be\label{pk}
\p = \frac{\k}{a_n}.
\ee
Starting from the vacuum, evolving through the window of times $-t_{pr}<t-t_n<t_{pr}$ where the $\chi$ particles reach their minimal mass $\tilde\mu$ generates a squeezed state
\be\label{squeezed}
|\Psi\rangle={\cal N} \exp\left(\int\!\frac{d^3 \k}{(2\pi)^3}\frac{\beta_{k}}{2\alpha_{k}^*} a_{\k}^\dagger a_{-\k}^\dagger\right)|0\rangle\,,
\ee
where ${\cal N}$ is a normalization factor, and the Bogoliubov coefficients satisfy
\be\label{beta0}
|\beta_{k}|^2=\exp\left(-\frac{\pi(\tilde\mu^2+\p^2)}{g\dot\phi}\right), ~~~~ |\alpha_{k}|^2-|\beta_{k}|^2=1.
\ee
This leads to a source for scalar (and tensor) perturbations  which is essentially a step function times a more slowly varying source
\be\label{source}
J=\frac12\chi^2\frac{\delta }{\delta\phi}m_\chi^2 \sim  n_\chi \frac{\delta m_\chi}{\delta\phi} \,,
\ee
where 
\be\label{nchi}
\langle n_\chi \rangle \equiv \bar n_\chi = \int \frac{d^3 \k}{(2\pi)^3 a_n^3} |\beta(k)|^2
\simeq (g\dot\phi)^{3/2}\exp\left(-\frac{\pi\tilde\mu^2}{g\dot\phi}\right)\,,
\ee
is the average number density of $\chi$ particles produced in each event. This density dilutes with the expansion of the universe, but every $2\pi/\omega$ a new generation of sources is produced. The physical momentum scale of the production events $p_\text{max}$ can be seen to be of order $(g\dot\phi)^{1/2}$ from (\ref{beta0}).   

\subsubsection{Bose enhancement and backreaction}

In certain situations we should consider the effect of previously produced particles on a given event. So far we discussed particle creation from the vacuum, but the calculation easily generalizes.  
If there are many production events per Hubble time, i.e. $ \omega/H\gg 1$, and if the massive particles of interest are bosons there will be an enhancement in their production. To apply the flat space analysis we restrict attention to a subset of the events occurring in one Hubble time. The number of produced particles in the presence of an existing number $n(k)$ excited is modified to
\be
\Delta n(k) = (1+n(k)) |\beta(k)|^2.
\ee
Using a continuum approximation\footnote{The continuum approximation is only for the purpose of estimating whether the effects of the previously produced particles can be neglected; in our main calculations we will of course keep track of the oscillations.} when there are many events with frequency $\omega$ and ignoring the redshift of momenta we obtain
\be
\dot n(k) = \omega |\beta(k)|^2 (1+n(k)).
\ee
In one Hubble time this gives
\be\label{buildup}
n(k\sim 0) = \exp\left(\frac{\omega}{H}|\beta(0)|^2 \right) -1.
\ee
In order to match current limits on oscillatory features in the primordial power spectrum \cite{PlanckInflation},  we will be interested in a regime with $\tilde\mu$ sufficiently large that $|\beta(0)|^2\sim e^{-\pi\tilde\mu^2/g\dot\phi}\sim 10^{-3}-10^{-2}$, and $\omega/H$ ranging up to of order $10^2$.  The effect of the previously produced particles therefore only becomes marginally important for the highest frequencies.  




Before moving to the perturbations, we should note the conditions for our produced $\chi$ particles not to strongly affect the background evolution of the inflaton $\phi$.  We can estimate the back reaction of effective scalar potential contributed by the $\phi$-dependent energy density in $\chi$ as $\rho_\chi\sim \sum g(\phi-\phi_n)n_\chi$.  First, we must keep its effect subdominant to the original slow roll background evolution by imposing
\be\label{smallrhochi}
g \bar n_\chi \ll V'(\phi)\sim 3 H\dot\phi\,.
\ee
In addition, we will impose 
\be\label{smallBR2}
\rho_\chi \sim m_\chi \bar n_\chi\ll M_P^2 \dot H\sim \dot\phi^2\,,
\ee
where in the last step we used that our background is slow-roll inflation.   This prevents the production of the $\chi$ particles from draining significant kinetic energy from the inflaton, $i.e.$ we are working far from the regime of \cite{trapped}.  
We will verify that these conditions are satisfied below in our parameter window of interest after deriving the perturbations.    

\subsubsection{Action for the fluctuations and their mode functions}

We will be interested in repeated production events whose distribution in time will determine the scale-dependence of our perturbations.  A discrete shift symmetry, as arises in axion monodromy inflation, will lead to shapes respecting a discrete version of scale-invariance, i.e. a symmetry under  $\log(k/H)\to \log(k/H)+2\pi n H/\omega$ for integer $n$.   The shapes will exhibit residual oscillatory features which we will compute in detail.

 
Working with the conformal time coordinate $\eta=-1/aH$, and decomposing $\phi(\eta,\x) = \phi_0(\eta)+\delta\phi(\eta,\x)$, the action is
\bea\label{Skinsource}
S &=& \int\!d^3\x d\eta \left\{ a^2(\eta) \frac12\left[(\partial_\eta\delta\phi)^2-(\partial_{\x}\delta\phi)^2
+(\partial_\eta\chi)^2-(\partial_{\x}\chi)^2-a^2(\eta) m_\chi^2(\phi_0(\eta))\chi^2\right] \right. \nonumber \\ 
&& ~~~~~ - \left.a^4(\eta)\left[ \delta\phi J+{\cal L}_4+{\cal L}_5+\dots\right]\vphantom{\frac12}\right\}. \nonumber \\
\eea
where 
\be\label{delL}
{\cal L}_j=\frac12\frac{1}{(j-2)!}\chi^2\delta\phi^{j-2} \frac{\delta^{j-2}}{\delta \phi^{j-2}} m_\chi^2
\ee
describes interactions higher order in $\delta\phi$ which descend from the $\phi$-dependent mass term.

The $N$-point functions can be computed by standard in-in perturbation theory
\be\label{inin}
\langle in|\overline{T}\left( e^{i\int_{-\infty(1+i\epsilon)}^t dt_1{\cal H}_{int}}\right)
\delta\phi_{k_1}(t)\dots\delta\phi_{k_N}(t)T \left(e^{-i\int_{-\infty(1-i\epsilon)}^t dt_2 {\cal H}_{int}}\right)
  |in\rangle \,,
\ee
where the interaction picture fields $\chi$ and $\delta\phi$ are evolved with the quadratic Hamiltonian, including the time-dependent mass-squared term for the $\chi$ particles, $\frac{1}{2}\chi^2 m_\chi(\phi_0(t))$ obtained from the background homogeneous evolution $\phi_0(t)$ of the inflaton. To evaluate~\eqref{inin} we need the mode functions for the scalar fluctuations and $\chi$ fields.\\

\noindent{\bf $\delta\phi$ mode function}\\[.2cm]
We start by expanding the interaction picture field in terms of lowering and raising operators $ a_\k$ and $ a_\k^\dagger$
\be
\delta\phi(\eta,\x)=\int_\k   a_\k u_k(\eta) e^{i\k\cdot \x}+h.c.\,,
\ee
which satisfy $[ a_\k,  a_{\k'}^\dagger]=(2\pi)^3\delta^3(\k-\k')$, as well as $ a_\k|in\rangle=0$,  
and we defined the shorthand notation
\be
\int_\k \equiv \int \frac{d^3\k}{(2\pi)^3}.
\ee
Considering the leading de Sitter expansion with approximately constant $H$, the properly normalized mode solution is
\be\label{umodes}
u_k(\eta)=\frac{H}{\sqrt{2 k^3}}(i-k\eta) e^{-ik\eta}\,.
\ee
These mode functions satisfy
\be\label{canonical}
a(\eta)^2 (u_k\partial_\eta u_k^*-u_k^*\partial_\eta u_k)=i\,,
\ee
ensuring canonical commutation relations for $\delta\phi$ and its canonical momentum $\Pi_{\delta\phi}=a^2(\eta) \partial_\eta \delta\phi$.
At early times this becomes
\be\label{early}
u\to -\frac{H\eta}{\sqrt{2k}}e^{-i k\eta}=\frac{1}{a^{3/2}\sqrt{2 (k/a)}}e^{-i k \eta}\,.
\ee

\noindent{\bf $\chi$ mode function}\\[.2cm]
Similarly for a given sector of $\chi$ particles, we have the mode expansion
\be\label{chigen}
\chi(\eta,\x)=\int_\k  a^{(in)}_{\chi, \k} v_k(\eta) e^{i\k\cdot \x}+h.c.\,,
\ee
where the mode function $v_k(\eta)$ is a solution of the free equation of motion including the effects of the time-dependent mass, and $a^{(in)}_{\chi, \k}|in\rangle =0$.   This encodes the evolution of the operator via the free Hamiltonian, appropriate for the interaction picture.   In case (b), with sinusoidal $\omega_\chi^2(t)$, inside the horizon this is a Mathieu function.   But there is a simple WKB approximation valid in our regime:  between bursts of particle production, the solution is a linear combination of adiabatic modes which we can write as
\be\label{vsoln}
a^{3/2} v_k(t)=\alpha_{k }^{(n)} \frac{\exp(- i \int_{t_n}^{t} dt' \omega_\chi(t'))}{\sqrt{2\omega_\chi (t')}}
+{\beta_{k }^{(n)}}^* \frac{ \exp(i\int_{t_n}^{t} dt' \omega_\chi(t'))}{\sqrt{2\omega_\chi (t')}}, ~~~~ t_n+t_{pr}<t<t_{n+1}-t_{pr}
\ee   
with normalization $|\alpha_{k}^{(n)}|^2-|\beta_{k}^{(n)}|^2=1$, where $t_{pr}$ is the timescale of the production event (\ref{tpr}).   We can consider a mode solution which is pure positive frequency initially, 
i.e. take $\alpha_{k}^{(0)}=1, \beta_{k}^{(0)}=0$.  After the first event, a nontrivial $\beta$ contribution is generated. 
The full set of ${\alpha_{k}^{(n)}, \beta_{k}^{(n)}}$ within the Minkowski regime (when Hubble dilution is negligible) can be understood in a simple way from the analogue Schrodinger problem solved by the mode solution
\be\label{vSchrod}
-\ddot v -\omega_\chi^2(t) v=0
\ee 
with effective potential $-\omega_\chi^2(t)$.  

In our specific case (a), at each time $t_n$, there is a different $\chi$ sector that reaches its minimal mass, so we have one production event per $\chi$ sector.  In case (b), we have a single $\chi$ sector with an oscillating mass leading to repeated particle production events for this sector.
The analysis leading to (\ref{buildup}) suggests that for our purposes, even in this latter case we can treat the events as independent, with the correlator being a sum over their contributions.       
In our case (a), for each sector of $\chi$ particles this is simply scattering off of an inverse Harmonic oscillator potential, with the reflection coefficient of order $|\beta|^2$ (see for example the appendix of \cite{SSZ} for a derivation).  In our case (b), this is scattering off a  sinusoidal potential, which behaves as a sequence of inverse Harmonic oscillators near its maxima.  As mentioned above, the full solution for this is given by Mathieu functions (one-dimensional Bloch waves in a sinusoidal potential).

The effect can be shuffled between the mode solution $v$ and the basis of creation and annihilation operators, via the Bogoliubov transformation $a_\chi^{(n+1)}=\alpha a_\chi^{(n)}+\beta {a_\chi^{(n)}}^\dagger$.  The state $|(n)\rangle$ satisfying $a_\chi^{(n)} |(n)\rangle =0$ is a squeezed state $\propto \exp\left(\int_\k\frac{\beta}{2\alpha^*}{a_{\chi\,\k}^{(n+1)}}^{\dagger} {a_{\chi\,-\k}^{(n+1)}}^{\dagger}\right)|(n+1)\rangle$ in terms of the ${a_\chi^{(n+1)}}$ Fock space with $a_\chi^{(n+1)}|(n+1)\rangle =0$.  


For a given particle production event at time $t_n$, we generate a squeezed state excited above the $|(n+1)\rangle$ vacuum.  To compute the contributions to the correlation functions from this event we can work with the Fock space built from ${a_{\chi}^{(n+1)}}^\dagger$.  (We will suppress the ${(n+1)}$ label in subsequent expressions.)  The formula corresponding to (\ref{inin}) is given by writing the state $|in\rangle$ in this basis,
\begin{multline}\label{ininsqueezed}
|{\cal N}|^2\langle (n+1)| e^{\int_q\frac{\beta_q^*}{2\alpha_q}a_{\chi,\q}a_{\chi,-\q}}
\overline T \left(e^{i\int_{-\infty(1+i\epsilon)}^t dt_1{\cal H}_{int}}\right)\times\\
\delta\phi_{k_1}(t)\dots\delta\phi_{k_N}(t)
\left(Te^{-i\int_{-\infty(1-i\epsilon)}^t dt_2 {\cal H}_{int}}\right)
e^{\int_k\frac{\beta_k}{2\alpha_k^*}a_{\chi,\k}^{\dagger}a_{\chi,-\k}^\dagger}  |(n+1)\rangle 
\end{multline}
with the normalization ${\cal N}=1+{\cal O}(|\beta|^2)$.  In our calculations below, we will find that the leading effects come from saddle points $t_{i*}$ in the integrals over $t_i$ in the interaction Hamiltonian, and that these saddles are at or after the production event, $t_{i*}\ge t_n$.  As a result, we can replace the lower limits of integration with $t_n$ to good approximation.  The expansion of $\chi$ in this basis is then simply of the form
\be\label{chiexp}
\chi(t, \k)= \ \frac{(H\eta)^{3/2} a_{\chi, \k} e^{-i\int_{t_n}^t dt' \omega_\chi(t')}}{\sqrt{2\omega_\chi(t)}} + h.c.\,.
\ee
The Bogoliubov coefficients describing the particle production now appear in the the state rather than the mode functions.  We have included the prefactor $(H\eta)^{3/2}$ appropriate for the de Sitter background.
We have written this in a WKB form, as is justified by the massiveness of the $\chi$ particles which leads to small variation $\dot\omega/\omega^2\ll 1$.  

As we will see, different classes of diagrams dominate in different regimes of parameters.   We will first consider the contributions generated by the 3-point vertex \mbox{${\cal L}_3=\delta\phi(\eta, \x) J(\eta)$}, focusing on those which do not involve annihilations of $\chi$ particles.  This will generate effects similar to those predicted by the classical model described in the appendix of \cite{MSSZ}.  The leading such contributions scale like the density of produced particles $|\beta|^2\sim \bar n_\chi$ times other factors determined by a simple stationary phase approximation to the time integrals.\footnote{In our templates for analysis we leave these in terms of the original time integrals.} 
In the following section, we will consider other diagrams which encode quantum interference and annihilation effects.  These include contributions scaling like $\beta\sim \bar n_\chi^{1/2}$, but with more rapidly oscillating integrands and hence different prefactors.  The two classes of diagrams dominate in different regimes of parameters which we will spell out below.  The shape of the non-Gaussianity in the first class we will consider is novel, whereas the second class of diagrams has more similarity to resonant non-Gaussianity \cite{resonantNG}.  



\subsection{Sourced contributions at order $|\beta|^2$}\label{sec:sourced}

We will begin by computing contributions that are similar to those that would arise from a classical source created by the production event.  These contributions to the correlators of the scalar fluctuations will be given as in \cite{MSSZ}\ by correlators of the source convolved with the retarded Green's function. 
The retarded Green's function derived from (\ref{umodes}) is
\be\label{green}
G(\eta,\eta')= i\theta(\eta-\eta') [u_k(\eta)u_k^*(\eta')-u_k^*(\eta)u_k(\eta')]\,,
\ee
so that
\be\label{phiGJ}
\delta\phi_k(\eta)=\int d\eta'  G_k(\eta, \eta') a^4(\eta') J_k(\eta')\,.
\ee
Here the factor of $a^4$ comes from the source term.  (As above, we defined $J$ itself in proper units.)
We will be interested in late-time observations, related to
\be\label{Gsimp}
G(0,\eta')=-\frac{H^2}{k^3}(\sin(k\eta')-k\eta'\cos(k\eta'))\equiv -\frac{H^2}{k^3}\hat g(k\eta').
\ee
Let us unpack the source (\ref{source}) a bit more.  We have 
\be\label{Jdef}
J(\x, t)= \frac12 \chi^2(\x,t) \frac{\delta}{\delta\phi}m_\chi^2(\phi)|_{\phi=\phi(t)}
\ee
where for now we are approximating $\phi$ by its background evolution.  Close enough to the event, this is of the form $J(\x, t)=g^2\phi(t) \chi^2(\x,t)$ for both cases (a) and (b), but we will require the later evolution of the source. This leads to
\bea\label{Jk}
J_{\k} &\sim& 
 \int_{\k'}  \frac{a^\dagger_{\chi \k^\prime}a_{\chi, \k-\k^\prime}}{m_\chi(\phi)}\frac{\delta}{\delta\phi}m_\chi^2(\phi)|_{\phi=\phi(t)}(H\eta)^3\nonumber \\[10pt]
&& ~~~
+\{ (a_\chi^\dagger)^2 + (a_\chi^2)\} ~~{\rm terms} 
\eea
where we replaced the $1/\omega_\chi$ from the product of $\chi$ mode functions with $1/m_\chi$. This is often a good approximation, for the following reason.    In both of our cases (a) and (b), the frequency can be written as
\be\label{fullomega}
\omega_\chi =\sqrt{\tilde\mu^2+\Delta m^2(t)+(k/a)^2}
\ee
where as above $\tilde\mu$ denotes the minimal $\chi$ mass, either $\mu_a$ or $\mu_b$, and $\Delta m(t)^2\ge 0$.  The $(k/a)^2$ term dilutes exponentially, and its initial value is the dominant momentum squared in our particle production process, $p_{\max}^2\sim g\dot\phi/\pi$.  This is smaller than $\tilde \mu^2$ in our regime of interest, for which the exponential Bogoliubov coefficient (\ref{expform}) is $\ll 1$.    
      
The second line in (\ref{Jk}) indicates the rest of the terms quadratic in raising and lowering operators.\footnote{Here we note that the $a^\dagger a$ term is equivalent to an $a a^\dagger$ term away from zero momentum; this is automatic if we define $\chi\chi$ by normal-ordering.}
These terms will lead to interesting interference effects we will include below, in fact starting at order $\beta$ (rather than order $|\beta|^2$ as we have here).  These contributions have a net oscillation with time $\simeq \exp(\pm 2 i \mu (t-t_n)$.  At a given order in $\beta$, this causes some suppression relative to effects we will find here which in our case (b) will resonate at the scale $k/a\sim \omega = \dot\phi/f$.  For the present section, we will therefore focus on the contributions that arise from the $a^\dagger a$ terms.    
The last factor captures the Hubble dilution of the source particles after their creation.   


The next step is to derive the correlators of (\ref{Jk}), from which using (\ref{phiGJ}) we will obtain the desired contribution to the correlators of $\delta\phi_k$.  
These are expectation values in the squeezed state
\be\label{squeezed}
|\Psi\rangle= {\cal N} \exp\left(\int_\k \frac{\beta_k}{2\alpha_k^*} a_{\chi, \k}^\dagger a_{\chi, -\k}^\dagger \right)
|0\rangle\,,
\ee
where ${\cal N}$ is a normalization factor, and 
\be\label{beta}
\beta_{\k}\simeq \exp\left(-\frac{\pi(\k^2/a_n^2+\mu^2)}{2g\dot\phi}\right)\tilde\theta((t-t_n)/t_{pr})\,.
\ee
Here $\tilde\theta$ is a step function smoothed out over the non-adiabaticity timescale (\ref{tpr}).
For the two point function we find the behavior
\be\label{Jtwo}
\langle \Psi | J_{\k_1}(\eta'_1)J_{\k_2}(\eta'_2) |\Psi \rangle \sim\\  { (2\pi)^3\delta(\k_1+\k_2)}\bar n_\chi
\prod_{j=1}^2\tilde\theta((t'_j-t_n)/t_{pr})2\frac{\delta}{\delta\phi} m_\chi(\phi(\eta'_j))(H\eta'_j)^3 \,.
\ee
Finally we can plug the above into (\ref{phiGJ}) to estimate the Gaussian scalar perturbations. 
In this step, we treat the $\tilde \theta$ functions as simply step functions $\theta(\eta'-\eta_n)$, since
the production timescale is much shorter than that of the oscillatory features we are considering.   We can view this as a test of the UV sensitivity of this part of the calculation -- if the result does not blow up, then evidently the high energy scale $1/t_{pr}$ scale is not cutting off any divergence.   
Defining
\be\label{hdef}
\hat h(k\eta_n)=\int_{\eta_n}^0 \frac{d\eta'}{\eta'}(\sin k\eta'-k\eta'\cos(k\eta'))\frac{\delta}{\delta\phi}m_\chi(\phi_0(\eta'))\,,
\ee
and using (\ref{phiGJ}) and (\ref{Jtwo}) we find the power spectrum
\bea\label{deltaphisq}
\langle\delta\phi_{\k_1}\delta\phi_{\k_2}\rangle_{pp} 
&\sim&   \frac{(2\pi)^3\delta(\k_1+\k_2)}{k_1^3}\left( \frac{\bar n_\chi}{H^3}\right)H^2 \sum_n\frac{\hat h(k_1\eta_n)^2}{ (-k\eta_n)^3}\,.
\eea
Translating to $\zeta$ using $\zeta\simeq - \frac{H}{\dot\phi}\delta\phi$ (plus slow-roll suppressed higher order corrections), and using the standard result
\be\label{zetavac}
\langle \zeta_{vac, \k_1}\zeta_{vac,\k_2}\rangle \sim \frac{H^4}{\dot\phi^2}\delta(\k_1+\k_2)\,,
\ee
we find
\bea
\expect{\zeta^2}_{pp}&\sim& \expect{\zeta^2}_{vac}\times  \frac{\bar n_\chi}{H^3} \sum_n\frac{\hat h(k_1\eta_n)^2}{ (-k_1\eta_n)^3}\,.
\eea
The subscript $pp$ refers to the particle production contribution.  The truly oscillatory contribution is a piece of this as we will describe below.  

For the N point function, following similar steps, we find a connected contribution
\be\label{Npoints}
\langle\delta\phi_{\k_1}\dots\delta\phi_{\k_N}\rangle\sim (2\pi)^3\delta\left(\sum \k_i\right)\frac{\bar n_\chi}{H^3}  H^{N+3}\sum_n(H\eta_n)^{-3}\prod_{i=1}^N \frac{\hat h(k_i\eta_n)}{k_i^3}\,,
\ee
which can similarly be traded for $\langle\zeta^N\rangle$.





In the regime where these contributions dominate, these formulas directly lead to templates for analysis, collected below in section \ref{templates}.  Let us examine their amplitude and shape.
One important quantity is the ratio of signal/noise in the three and two point functions, which in the cosmic variance dominated Gaussian approximation is given by
\be\label{SNratio}
\frac{(S/N)_3}{(S/N)_2}\sim \frac{\expect{\zeta^3}'_{pp}/(\expect{\zeta^2}'_{vac})^{3/2}}
{\expect{\zeta^2}'_{pp}/\expect{\zeta^2}'_{vac}}\sim \frac{\sum_n (k\eta_n)^{-3} \hat h(k\eta_n)^3}{\sum_{n'} (k\eta_{n'})^{-3} \hat h(k\eta_{n'})^2}\,,
\ee
where the prime on expectation values denotes dropping $(2\pi)^3\delta^3(\sum_i \k_i)$ and we evaluated the numerator and denominator at $k_i=k$ to get a sense of the relative amplitudes; we will discuss the shape in momentum space (including the scale dependence) further below.  





To proceed, let us apply this result to a situation with an approximate discrete shift symmetry, with events evenly spaced in proper time $t$, corresponding to conformal times
\be\label{etan0}
\eta_n=-\frac{1}{H}e^{2\pi \frac{H}{\omega}(n+\frac{\gamma}{2\pi})}.
\ee
where $\omega = \frac{\dot\phi}{f}$ depends inversely on the underlying field period $2\pi f$.    In our case (b), this frequency appears in the cosine term in the potential.  In both cases (a) and (b) it describes the frequency of particle production events:  $\omega/2\pi H$ events per Hubble time.

The behavior of these $N$-spectra, and their ratios, is somewhat different in our two cases (a) and (b).    
Case (a) will prove to overlap strongly with existing templates for $\omega/H\ge 1$, whereas case (b) has additional resonances in the time integrals as a result of the oscillating mass and has small overlap with existing templates for $\omega/H\ge 1$.   So in much of this work we will focus on the behavior of case (b).  But let us evaluate them in turn.

\subsubsection{Estimates for the integrals in case (a)}

In this case, from (\ref{2amasses}) we have approximately a step function source since
\be\label{delma}
\frac{\delta m_\chi}{\delta \phi}\to g
\ee
for $t-t_n > t_{pp}$, equivalently $\phi-\phi_n > \sqrt{2}\mu/g$.  Thus the integral we need is
\be\label{h}
\hat h(k\eta_n) = g\int_{\eta_n}^0 \frac{d\eta'}{\eta'}(\sin k\eta'-k\eta' \cos k \eta').
\ee
The sum in (\ref{deltaphisq}) and (\ref{Npoints}) is dominated at the horizon crossing time because the summand becomes small if for any of the momenta $-k_i\eta_n$ is much different from 1. When $-k_i\eta_n\ll 1$ the Green's function is suppressed as $(k_i\eta)^3$.
When $-k_i\eta_n\gg 1$ we have
\be
\hat h(k_i\eta_n)/g \simeq \frac{1}{2} {\rm Im}\ {\rm Pr.}\int_{-\infty}^\infty \frac{d\eta'}{\eta'} e^{ik_i\eta'}
-k_i\int_{\eta_n}^0 d\eta' \cos k_i\eta' = \frac{\pi}{2}+\sin k_i\eta_n = \O(1).
\ee
This justifies glossing over the short scale details of the production event and approximating the source by a step function.
Incorporating this, we estimate the relevant ratio for $S/N$ in the $N$-point function (\ref{Npoints}) as
\be\label{abetasquared}
\frac{\expect{\zeta^N}_{pp}}{\zeta_{\rm vac}^N} \sim  N_X g^N.
\ee
where
\be\label{NX}
N_X=\frac{\bar n_\chi}{H^3}\frac{\omega}{H}
\ee
is the number of events per Hubble time and Hubble volume.  

\subsubsection{Estimates for the integrals for case (b)}

We would like to estimate the dominant contributions to the integral over $\eta'$:
\be\label{hbint}
\hat h_b(k\eta_n)=c_b\int_{\eta_n}^0 \frac{d\eta'}{\eta'} \hat g(k\eta')\left.\frac{\delta m_\chi}{\delta\phi}\right|_{\phi=f \omega t=f\frac{\omega}{H}\log\eta' H} \,.
\ee
From (\ref{bmasssquaredgf}) we have
\be\label{massexpand}
\frac{\delta m_\chi}{\delta\phi} = -\frac{(g^2 f/\mu) \sin\frac{\phi}{f}}{\sqrt{1+\frac{2 g^2 f^2}{\mu^2} \cos\frac{\phi}{f}}}\,.
\ee
For sufficiently small $g^2 f^2/\mu^2$, this reduces to the simpler form $g^2\frac{ f}{\mu} \sin\frac{\phi}{f} $, giving the integral
\be\label{hbint}
\hat h_b(k\eta_n)=c_b\int_{\eta_n}^0 \frac{d\eta'}{\eta'} \sin (\frac{\omega}{H}\log\frac{\eta'}{\eta_n})\ \hat g(k\eta')\,,
\ee
with coefficient
\be\label{cb}
c_b\sim g^2\frac{f}{\mu}
\ee
from (\ref{bmasssquaredgf}).   

We will present our final analysis of the parameter windows for our template based on the approximation (\ref{hbint}) below in \S\ref{sec:parameterwindows}.   We will find there that for the lower end of frequencies $\omega$ of interest, the ratio $2 g^2 f^2/\mu^2$ is not hierarchically suppressed.  We should therefore either include the full form (\ref{massexpand}) (along with an extra parameter $2 g^2 f^2/\mu^2$ varying over a small range of values), or determine that the overlap between the two templates is strong enough to justify the simplification to the pure sinusoidal function. 
To begin, we will analyze the pure sinusoidal form in case (b), and then return to this point.  By calculating the Fourier coefficients of the full expression, we find that even for $2 g^2 f^2/\mu^2\lesssim 1$, the simpler expression gives very similar results for the leading Fourier mode contributing to the resonant integral.


Let us now estimate the size of this effect by approximating the integral over $\eta'$ and sum over $n$. First, note that the Green's function $\hat g(k\eta')$ (\ref{Gsimp}) is of order $(k\eta')^3$ as $k\eta'\to 0$, suppressing any contributions outside the horizon.  For $-k\eta'>1$, the second term in $\hat g(k\eta')$ dominates over the first.   The dominant contribution to the integral in (\ref{hdef}) is easily estimated by a stationary phase approximation, taking into account the two sources of oscillation in the integrand in case (b).  That is, for $k\eta' >1$ the integrand has two oscillating functions: $g(k\eta')\sim k\eta' \cos k\eta'$ and $\sin \frac{\omega}{H}\log(\eta'/\eta_n)$, which resonate at $-k\eta' = \frac{\omega}{H}$.  Explicitly,
\be\label{phaseint}
\hat h_b(k\eta_n) =-\frac{i}{4}k c_b\int_{\eta_n}^0 d\eta e^{i k\eta+i\frac{\omega}{H}\log(\eta/\eta_n)} + \text{c.c.} + \text{non-resonant.}
\ee
A saddle point integration with $k\eta_{saddle}=-\alpha$ leads to
\be\label{hasymp1}
\hat h_b(k\eta_n) \sim c_b\sqrt{\frac{\omega}{H}} \cos\left(\frac{\omega}{H} \log (-k\eta_n)+\gamma
\right).
\ee
valid for $-k\eta_n \gg \frac{\omega}{H}$ (so that the saddle point is well separated from the endpoint of the integral).   The leading contributions to the sum in both numerator and denominator of \eqref{SNratio} then come from the smallest value of $-k\eta_n$ which is consistent with picking up this saddle, i.e. $-k\eta_n\sim\alpha$.  From (\ref{etan0}) we note that of order $\alpha\equiv \omega/H$ terms in the sum over $n$ contribute with approximately the same value of $\eta_n$ (there are $\omega/2\pi H$ events within a Hubble time $H^{-1}$).   
Altogether this leads to a ratio (\ref{SNratio}) of order 
\be\label{SNR}
\frac{(S/N)_3}{(S/N)_2}\sim c_b\sqrt{\alpha}=c_b\sqrt{\frac{\omega}{H}}\,.
\ee
This behavior can be checked numerically.  This result also controls $(S/N)_{N+1}/(S/N)_N$ at tree level, within a finite range of $N$ for which the cosmic variance limited Gaussian approximation applies.   We will find in \S\ref{sec:parameterwindows}\ that (\ref{SNR}) can be somewhat larger than 1 consistently with our conditions for control of the model.   

Let us also separately record the amplitude of the correction to the power spectrum and bispectrum.  
For the power spectrum, from (\ref{deltaphisq}) we obtain
\be\label{pppower}
(S/N)_2 
\sim 
\frac{\expect{\zeta^2}_{pp}}{\zeta_{vac}^2}\sim \frac{\bar n_\chi}{H^3} \frac{c_b^2 }{\alpha}\,.
\ee
Here the last factor comes from a product $\alpha\times 1/\alpha^3\times (c_b\sqrt{\alpha})^2$, which arises respectively from the presence of of order $\alpha$ terms contributing to the leading terms of the sum with $-k\eta_n\sim\alpha$, a factor of $1/\alpha^3$ from the $1/(k\eta_n)^3$, and finally a factor $c_b^2(\sqrt{\alpha})^2$ from the above saddle point estimate for $\hat h$, appearing quadratically in the two-point function.

The sensitivity is bounded by $1/\sqrt{N_{\rm modes}}\sim 10^{-3}$, and in practice the data constrains this ratio in a frequency-dependent manner, ranging from $\sim 0.2$ at the highest frequency within the controlled effective field theory regime, $\omega\sim f$ to of order $10^{-2}$ at lower frequencies (see e.g. figures 37-38 of \cite{PlanckInflation}).  

Expressed in terms of an $f_{NL}$ parameter, the amplitude of this contribution to the bispectrum is
\be\label{fNL}
f_{NL}^{(b)}\equiv k^6 \frac{B(k, k, k)}{4 P_\zeta^2}\sim \frac{\bar n_\chi}{H^3} \frac{1}{2}\frac{\dot\phi}{H^2}\sum_n \frac{\hat h(k\eta_n)^3}{(k\eta_n)^3}\sim \sqrt{\frac{1}{P_\zeta}}c_b^3\frac{\bar n_\chi}{H^3} \times \sqrt{\frac {H}{\omega}}\,,
\ee
where
\be\label{Bdef}
\langle \zeta_{k_1}\zeta_{k_2}\zeta_{k_3}  \rangle = (2\pi)^3\delta^3(\k_1+\k_2+\k_3) B(k_1, k_2, k_3) \,,
\ee
and $P_\zeta\equiv H^4/(2\dot\phi^2)$.  In the last step, we estimate the size of the sum as 
described above,
\be\label{sumexplanation}
\sum_n \frac{\hat h(k\eta_n)^3}{(k\eta_n)^3}\sim  \alpha\times \frac{1}{\alpha^3}\times (\sqrt{\alpha})^3\times c_b^3 =c_b^3\alpha^{-1/2}.
\ee


\subsubsection{Fourier coefficients of full mass formula}

We will describe the controlled parameter windows in which this result pertains in the next section \S\ref{sec:parameterwindows}, after treating other contributions to our correlators.  Now let us return to the question raised above regarding the validity of simplifying the mass formula (\ref{massexpand}) neglecting the second term in the denominator.  This is clear for small 
\be\label{kappa}
\kappa \equiv \frac{2 g^2 f^2}{\mu^2}.
\ee
For $\kappa \lesssim {\cal O}(1)$, the Fourier transform of the full expression will have potentially significant contributions from higher harmonics as well as additional contributions to the original $e^{\pm i\omega t}$ terms.
\be\label{massexpand}
\frac{\delta}{\delta\phi} m_\chi = -\frac{(g^2 f/\mu) \sin\omega t}{\sqrt{1+\frac{2 g^2 f^2}{\mu^2} \cos\omega t}}=-\frac{g^2 f}{\mu}\left(F_1 e^{i\omega t}+F_{-1}e^{-i\omega t}+F_2 e^{2i\omega t}+F_{-2}e^{-2i\omega t}+\dots\right)
\ee
where the Fourier coefficients $F_j$ depend on $\kappa$.
The higher harmonics have a larger effective frequency, $\omega_{eff}/H\sim N\omega/H$.   This will be suppressed when integrated against our Green's function and summed over events, since as we have seen, the net results (\ref{pppower}) and (\ref{sumexplanation}) are proportional to a negative power of $\omega_{eff}/H$.     

Numerically computing the Fourier coefficients even for $\kappa$ close to 1 leads to similar results to those we obtain from the truncated expression without the square root denominator.  
For $\kappa=0.999$ for example, the coefficient of the $e^{\pm i\omega t}$ term in $\sin(\omega t)/\sqrt{1+0.999\cos(\omega t)}$ is $\sim \pm0.6 i $, which is close to the original coefficient $\pm 0.5 i$ without the square root factor.    The coefficient of the constant term is zero.  The higher Fourier coefficients $e^{\pm i N\omega t}$ are $\sim 0.2, 0.15, 0.11$ for $N=2, 3, 4$ respectively.  As mentioned above, when included in our full calculation, these are also somewhat further suppressed by our resonant integral and sum, as can be seen directly from our previous results with the replacement $\omega/H\to N\omega/H$.  Although they are somewhat suppressed, it may be worthwhile to include a small number of the additional harmonics in comparing the theory to data.  We will comment on this further in laying out the templates for analysis below in section \S\ref{templates}.      

\subsection{Contributions from higher point vertices at order $|\beta|^2$}\label{higherbetasquared}

Let us next consider the contributions from higher point vertices in the interaction Lagrangian.

\subsubsection{Case (a)}

At two points, we can have one of the members of a pair of produced particles emit a pair of $\delta\phi$ perturbations via an insertion of $\int a^4 {\cal L}_4$ with ${\cal L}_4\sim g^2 \chi^2\delta\phi^2$ (whereas two insertions of the time-dependent three-point interaction ${\cal L}_3\sim g^2(\phi(t)-\phi_n) \chi^2\delta\phi $ entered into the above contribution).   We find the ratio of this contribution to the one we computed in the previous subsection can be estimated as follows. First note that when considering two ${\cal L}_3$ there is an extra pair of $\chi$ fields which lead to a factor of $1/m_\chi(t)$ from their mode functions. We saw that in case (a) the leading effect comes from emission of soft quanta with frequency of order $H$, so we can take the production event to be instantaneous. For $t\gg t_{pr}$ given in \eqref{tpr} we can approximate $m_\chi = g(\phi- \phi_n)$. Thus the new contribution compared to the old one is
\be\label{ratioa}
\frac{H}{g (\phi-\phi_n)} \le \frac{H}{g\dot\phi t_{pr}}\sim \frac{H}{\mu_a} \ll 1
\ee

\subsubsection{Case (b)}
Similarly for case (b), at two points we can have one of the members of a pair of produced particles emit a pair of $\delta\phi$ perturbations via an insertion of $\int a^4 {\cal L}_4$.  The ratio of this contribution to the one we computed in the previous subsection is of order
\be\label{ratio}
\frac{\mu H}{g^2 f^2}\sqrt{\frac{\omega}{\pi H}}=\frac{\mu^2}{g^2 f^2}\times \frac{H}{\mu}\times\sqrt{\frac{\omega}{\pi H}}>\frac{H}{\mu}\sqrt{\frac{\omega}{\pi H}} 
\ee
The first factor here follows from the fact that each additional $\chi\chi$ from an interaction introduces a factor of $H/\mu$, and the factor of $\sqrt{\omega/\pi H}$ arises from the  structure of the power law contributions to the resonant time integrals:  at a resonance, each measure factor $d \eta$ gives $\sqrt{\pi (\omega/H)}/k$, while each factor of $\eta$ evaluated on the saddle gives a factor of $-(\omega/H)/k$.  For higher point vertices, there are fewer $d\eta$ measure factors; the same scaling with $k$ arises from a compensating number of factors of $\eta$.   
The last inequality follows simply from the fact that $g f <\mu$ in order for the $\chi$ mass-squared to remain positive throughout the process.  Altogether, since $H/\mu \ll 1$, there is a window in which the previous contribution dominates, including a range with $\omega>H$.    
Similar comments apply to the higher $N$-point functions, for which diagrams with higher point vertices (up to an $N+2$ point vertex) contribute.  

The shapes depend on the diagram topology in the following way.
The diagram generated by the highest point vertex has a shape similar to resonant non-Gaussianity \cite{resonantNG}, with the sinusoidal part of its scale-dependence entering as a function of the total momentum $k_1+k_2+\dots k_N$.  This is in contrast to the product structure we found in \S\ref{sec:sourced}, from the diagrams  with only 3-point vertices.  

Let us spell this out a little more explicitly for the bispectrum, which is schematically of the form
\bea\label{bibetasquared}
& & \frac{A}{k_1^2k_2^2k_3^2}\sum_{n=n_{min}}^\infty \left(\prod_{J=1}^3\frac{1}{-\eta_n k_J}\right)\left\{\prod_{I=1}^{3}\cos\left(\tilde\gamma_I + \frac{\omega}{H}\log(-k_I\eta_n)\right)\right. \nonumber \\
& &+ C_{34}\frac{k_2 k_3}{(k_2+k_3)^2} \cos\left(\gamma_{34}+\frac{\omega}{H}\log(-(k_2+k_3)\eta_n)\right)\cos\left(\tilde\gamma_{34} + \frac{\omega}{H}\log(-k_1\eta_n)\right)+{\rm permutations}\nonumber \\
& & \left.+ C_5 \frac{k_1k_2k_3}{k_T^3} \cos\left(\gamma_{5}+\frac{\omega}{H}\log(-(k_1+k_2+k_3)\eta_n)\right) \vphantom{\prod_{I=1}^3}\right\}\,.
\eea
The top line is what we computed above, the contribution from three insertions of the 3-point interaction ${\cal L}_3$; its amplitude $A$ is given above in (\ref{shapesaddle}).  The next line contains the contribution from one insertion of ${\cal L}_3$ and one of ${\cal L}_4$.
The last  line contains the contribution from a single insertion of ${\cal L}_5$ and $k_T \equiv k_1+k_2+k_3$.  
Again, there are regimes where either the first line or the last dominates, and an intermediate regime where they all contribute.  
        
\subsection{Interference terms at order $\beta$}        \label{sec:interference}

{\it Wherever a part of society possesses a monopoly of the means of production, the worker, free or unfree, must add to the labor necessary for his own maintenance an extra quality of labor in order to produce the means of subsistence for the owner of the means of production.}  $~~~~~$ --Karl Marx
        
The interaction vertices $ {\cal L}_N$ (\ref{delL}) contain the operator $\chi^2(t')$, whose oscillator expansion includes terms quadratic in raising operators $\sim a_\chi^{\dagger 2} e^{2i\int_{t_n}^t dt' \omega_\chi(t')}$ or in lowering operators $\sim a_\chi^2 e^{-2i\int_{t_n}^t dt' \omega_\chi(t')}$ in addition to the $a_\chi^\dagger a_\chi $ terms discussed above.   This leads to additional contributions, including processes in which the produced $\chi$ particles annihilate.  

This leads to two qualitative effects on the resulting correlators:

$\bullet$ The leading contribution arises at order $\beta$, since the net number of creation or annihilation operators can annihilate against a term $\beta a_\chi^{\dagger 2}|0\rangle$ from the squeezed state.
This leads to an enhancement factor $\sim e^{\pi\tilde\mu^2/2g\dot\phi}$ in the amplitude relative to the contributions of order $|\beta|^2$.

$\bullet$ These terms come with an extra oscillation in the integrand.  In case (a), this leads to suppression factors. In case (b), the oscillation is  $\simeq e^{\mp 2 i \mu (t'-t_n)+\dots}$.  For $\mu>\omega$ this dominates over the oscillation from the $\cos(\phi/f)\to e^{\pm i\omega(t-t_n)}$, and in that regime it also leads to some power law suppression factors.  

It is a detailed question whether these contributions dominate over the order $|\beta|^2$ contributions computed above, given the numerical values of our parameters.  In the regime relevant to oscillatory feature searches, the exponential $\exp(\pi\tilde\mu^2/2g\dot\phi)$ can compete against power law factors. We find a regime of parameters where the power law suppression factors overcome the exponential enhancement, and others that go the other way.

The detailed shape again depends on the distribution of $n$-point vertices in the diagram. 
Starting from the general expression (\ref{ininsqueezed}), we can write the correlator at order $\beta$ as
\be\label{firsttimp}
\langle 0|\overline T\left(e^{i\int_{-\infty(1+i\epsilon)}^t dt_1{\cal H}_{int}}\right)\delta\phi_{k_1}(t)\dots\delta\phi_{k_N}(t)T\left(e^{-i\int_{-\infty(1-i\epsilon)}^t dt_2 {\cal H}_{int}}\right){\int_\k \frac{\beta_k}{2\alpha_k^*}a_\k^\dagger a_{-\k}^\dagger}  |0\rangle  + c.c.
\ee
Each interaction vertex has the operator $\chi^2$. The only non-vanishing contributions we can get come from terms in which at least one of these contains two lowering operators to absorb the created pair from the squeezed state.   The rest must have the same net number of creation and annihilation operators.    However, if we consider a contribution with some interaction vertices introducing $a_\chi^2e^{-2i\mu t}$ and others $a_\chi^{\dagger 2}e^{2i\mu t}$, they do not both resonate.\footnote{This is true regardless of which exponential $\exp(\pm i\int {\cal H}_{int})$ each vertex comes from.  If it comes from the anti-time ordered exponential, it must contain $a_\chi^{ 2}e^{-2i\mu t}$ and also the lowering operator for the $\delta\phi$ perturbation, so altogether $\exp(+2i(\mu/H) \log(\eta)-ik\eta)$ which does not resonate.}    

So we can focus on the $a_\chi^\dagger a_\chi$ contributions from the remaining $\chi^2$ operators.   For these contributions, note that we could get resonant contributions from both time-evolution operators, since the sinusoidal time-dependent background $(\d^n/{\d\phi^n})m_\chi^2$ contains both types of terms $e^{\pm i\omega t}$.   

Now let us consider the distribution of interaction vertices.
One extreme case is where we bring down a single $(N+2)$-point interaction vertex 
\be
\int a^4\chi^2\delta\phi^N (\d^n/{\d\phi^n})m_\chi^2,
\ee
whose $\delta\phi^N$ contracts with the $\delta\phi$'s in our $N$-point function.  
In fact we find that for $\mu>\omega$ there is no other contribution for which all of the time integrals extend through their resonant saddle points.  This can be seen as follows.    
First, we note that the interaction term we bring down that contains the $a_\chi^{ 2}e^{-2i\mu t}$ must be to the left of all the others in order to generate a connected contribution.  We can write the correlator as a commutator of $a_\chi ^2$ with the operators to its right, and this generates a separate delta function constraint on the sum of $\delta\phi$ momenta in those operators.  If this interaction comes from the anti-time ordered evolution operator on the left, it multiplies the term in $\delta\phi$ with an annihilation operator, $\propto e^{-ik\eta}$ and there is no resonance.  On the other hand, if this interaction comes from the time-ordered evolution operator on the right, its time integral ranges over later times than those of the operators to its right.  Those operators resonate at the frequency $k\eta_*\sim -\omega/H$, which is a later time than the resonance at $k\eta_*\sim -2\mu/H$.  So we cannot obtain both resonances within our integral.


Let us therefore consider the single insertion of ${\cal L}_{N+2}$.   For case (a), the only such vertex arises for $N=2$, and hence in that case these resonant contributions can only arise for the power spectrum.  For this reason, we will focus first on case (b) and then comment about the power spectrum contribution in case (a).  
The $a_\chi^2 e^{-2i\mu t}$  factor contracts with the pair of particles in our expansion of the squeezed state, with the corresponding momentum $\k$ integrated against the Bogoliubov coefficient $\beta$, which is Gaussian.  
There is momentum dependence in the full mode functions for $\chi$ (\ref{chiexp}).   We will need to include the first sub-leading piece in the expansion
\be\label{omegachi}
\omega_\chi = \mu +\frac{\k^2}{2\mu a^2}+\dots
\ee
so the mode solution for $\chi$ becomes
\be\label{chimode2}
\chi(k\eta) \simeq \frac{(H\eta)^{3/2}}{(2\sqrt{k^2/a^2+\mu^2})^{1/2}} e^{-i\int^t \sqrt{k^2/a^2+\mu^2} d\tilde t}
\ee
For particles that are produced at $t_n$ this becomes
\be
\chi(k\eta)\simeq \frac{(H\eta)^{3/2}}{(2\sqrt{\mu^2})^{1/2}} e^{-i\mu(t-t_n)}e^{-i\frac{k^2}{4 \mu Ha^2}((a/a_n)^2 -1)}.
\ee
Then the integral over $\k$ is approximately 
\be\label{pint}
\int_\k \beta(k) e^{-i\frac{k^2}{4 \mu Ha^2}((a/a_n)^2 -1)}
\simeq 2^{3/2} e^{\pi\mu^2/2g\dot\phi} a_n^3 \bar n_\chi
\left(1-i\frac{g\dot\phi}{\pi\mu H}\left((a/a_n)^2 -1\right)\right)^{-3/2}
\ee
which can give us an additional suppression for times such that the second term in the parentheses dominates.
This depends on the dominant (resonant) contribution to the integral over time, which as we will describe shortly is given by $k\eta_*\sim -2\mu/H$, and we get the dominant terms in the sum to be at $k\eta_n$ of this order. As a result $(a_*/a_n)^2-1 = \O(1)$.  The significance of this correction thus depends on the size of $g\dot\phi/\pi\mu H$, which can indeed be larger than one in part of our parameter space.

Putting this together, we can write the time integral going into this correlator as
\be\label{rightN}
\int_{\eta_n}^{0}\frac{d\eta_1}{\eta_1}\prod_{I=1}^N(-ik_I\eta_1 e^{ik_I\eta_1})e^{-i\frac{2\mu\pm \omega}{H}\log(\frac{\eta_1}{\eta_n})}\left(1-i \frac{g\dot\phi}{\pi\mu H}\left[\frac{\eta_n^2}{\eta_1^2}-1\right] \right)^{-3/2} + c.c.
\ee
The resonance is at
\be\label{resvalues}
k_T\eta_{1*} =-\frac{2\mu\pm\omega}{H}, ~~~k_T=k_1+k_2+\dots k_N\,.
\ee
This contribution to $\langle \delta\phi_{k_1}\dots\delta\phi_{k_N}\rangle$ generated by the $N+2$-point vertex is therefore approximately given by
\bea\label{shapesaddle}
& & \pi^{1/2}2^{-3}H^N\frac{\bar n_\chi}{H^3}e^{\frac{\pi\mu^2}{2g\dot\phi}} \frac{(2\mu\pm\omega)^{N-1/2}}{H^{N-1/2}}(\frac{g^2 f^2}{2\mu H})\frac{H^N}{f^N} 
\left(1-i \frac{g\dot\phi}{\pi\mu H}\left[\frac{\eta_{n_{min}}^2}{\eta_{1*}^2}-1\right] \right)^{-3/2}\nonumber \\ 
& &  \times \frac{1}{k_T^{N-3}}\prod_{I=1}^N\frac{1}{k_I^2}
 \sum_{n=n_{min}}^\infty (-k_T\eta_n)^{-3}\cos\left(\gamma_N+\frac{2\mu\pm\omega}{H}\log(-k_T\eta_n)\right) \,, 
\eea 
where the sum on $n$ starts at the minimal $n$ such that $-k_T\eta_{n_{min}}> -k_T\eta_{I*}$.  This term dominates in the sum, and introduces a suppression factor of order $(H/[2\mu\pm \omega])^3$.     

Again, here we are considering the regime $\omega<\mu$. In the opposite case, the resonances are all at $\omega$ and the situation is similar to the order $|\beta|^2$ contributions discussed above.  In that case, the contribution generated by the $N+2$-point vertex is similar to the resonant shape, oscillating like $\cos(\gamma+\frac{\omega}{H}\log(k_T))$.  

Finally, let us return to case (a), where as mentioned above this correction only affects the power spectrum ($N=2$).
In this case, the resonant integral is of the schematic form
\be\label{aintbeta}
\int d\eta' e^{i k_T\eta'} e^{-i(t'-t_n)^2g\dot\phi}=\int d\eta' e^{i k_T\eta' - \frac{g\dot\phi}{H^2}(\log\frac{\eta'}{\eta_n})^2}\,,
\ee
giving a saddle point equation
\be\label{saddleabeta}
k_T\eta'_*\sim 2\frac{g\dot\phi}{H^2}\log\left(\frac{k_T\eta'_*}{k_T\eta_n}\right)\,.
\ee
Defining $z=\eta_*/\eta_n < 1$, this is
\be\label{zsaddle}
z = -2\frac{g\dot\phi}{H^2(-k_T\eta_n)}\log(z)\,.
\ee
When the coefficient on the right hand side is large in magnitude, as it is for $-k_T\eta_n$ of order 1, the solution is $z\simeq 1$, i.e. $k_T\eta_*\approx k_T\eta_n$.  For larger $-k_T\eta_n$ the solution for $z$ decreases.  In the latter regime, the $(k_T\eta)_n^{-3}$ in the analogue of (\ref{shapesaddle}) for case (a) will suppress the integral.  
So we can focus on the former case.  In that case, we have a dominant contribution to the integral over $\eta'$ and sum over $n$ with $k_T\eta'_*\sim k_T\eta_n$ of order 1.  

Given this, the factors from the resonant integral are all of order one.   We then get an estimate for the size of this order $\beta$ contribution to the power spectrum in case (a) to be
\be\label{betaa}
\frac{\langle \zeta \zeta \rangle_{pp}}{\zeta_{vac}^2}\sim g^2 \frac{\bar n_\chi}{H^3} \frac{\omega}{H}  e^{\frac{\pi\mu^2}{2g\dot\phi}} \frac{H}{g(\phi-\phi_n)|_{t_*}}\,,
\ee
with the last two factors giving the ratio between this and the order $|\beta|^2$ contribution above in (\ref{abetasquared}).  
These factors arise as follows.  The exponential enhancement is due to the fact that this is of order $\beta$ not $\beta^2$.  The power law suppression in the last term arises from a combination of the ratio of couplings (four point versus three point) and the $1/m_\chi$ that comes with each vertex from its $\chi^2$ factor.

\subsection{Fermion production}

We have analyzed in detail the perturbations generated by the interaction between $\chi$ and the scalar $\delta\phi$.
In this section, let us briefly comment on the contribution of fermion production to the scalar perturbations.  This is interesting as a general possibility, and more specifically is relevant to our effect since the radiative stability under loop corrections was guaranteed above by microscopic supersymmetry.  

Fermion particle production was studied in detail in the second reference in \cite{otherproduction}, leading to results similar to the bosonic case for the Bogoliubov coefficients and average number density (cf equations (28)-(29) there).  To compare to our current analysis, consider the form of the interaction Hamiltonian descending from the fermion action, which, inside the horizon, reads
\be\label{Faction}
\int d\eta d^3\x a(\eta)^4\left\{ i\bar\psi_\chi \partial_\mu\gamma_\mu \psi_\chi - \bar\psi_\chi \psi_\chi\sqrt{\mu^2+g^2f^2\cos\frac{\phi}{f}} \right\}
\ee     
This leads to similar $j$-point interactions as we had for the bosons $\chi$ in (\ref{Skinsource}) and (\ref{delL}), 
\be\label{LNFermion}
{\cal L}_j=\frac{1}{(j-2)!}\bar\psi_\chi\psi_\chi \delta\phi^{j-2} \frac{\delta^{j-2}}{\delta \phi^{j-2}} m_\chi
\ee
Noting that $\langle\bar\psi \psi\rangle\sim \bar n\sim \langle\chi^2\rangle \mu$, we see that the parametric scaling of the fermionic and bosonic diagrams will be similar.  There will be some differences in detail in the function of $\sin\omega t$ and $\cos\omega t$ that appears at each order, depending on the ratio $gf/\mu \lesssim 1$.  But these coefficients remain approximately sinusoidal with frequency $\omega$, resonating at the same scale.    
Their amplitude and shape -- including the interpolation between a product structure and the resonant shape -- is
similar to that of the bosons.  We therefore do not expect the fermion contribution to significantly change the pattern of perturbations.  However, it would be interesting to investigate this further and determine whether ultimately fermions and bosons could be distinguished
in the event of a detection of this general class of shapes.  

\section{Parameter windows and amplitude comparisons}\label{sec:parameterwindows}

Next let us compare the amplitudes of the different classes of contributions worked out in the previous section.   We need to make two types of comparisons.  We must determine when different contributions to a given $N$-point function dominate, and in a given regime we would then like to determine the relative signal/noise in the non-Gaussianity as compared to that of the oscillatory features in the power spectrum.

The diagrams with a single higher-point vertex are similar to the resonant shape.  Such searches are not new, being motivated by the dynamics \cite{monodromy}\cite{resonantEFT}.  But the amplitude and the corresponding ratio $(S/N)_3/(S/N)_2$ is of interest in determining the level of motivation for joint power and bispectrum analyses for this shape. This ratio is less than unity for resonant non-Gaussianity \cite{resonantEFT}.

The diagrams generated from the 3-point vertices lead to the novel shape derived above.  In this case, we need to know the regime in which it dominates in order to prescribe parameter ranges for data searches motivated by this mechanism.      

The hybrid case, where insertions of all possible $j$-point vertices (up to $j=N+2$) compete, gives a rich shape with some overlap with the resonant shape, but extra structure in the additional terms.  In this case, the coefficients must be of the same order, which implies a restricted (and hence more predictive) regime of parameters for this mixed shape.  

\subsection{Controlled parameter window with leading contribution from ${\cal O}(|\beta|^2)$ sourced product shape}

Let us first assess the parameter window in which the order $|\beta|^2$ contributions computed in \S\ref{sec:sourced}\ dominate, with the theory satisfying all our conditions for self-consistency of the calculation.

In case (a), we will focus on the regime $\omega/H\le {\cal O}(1)$ where the shape requires a new search according to appropriate overlap calculations, as discussed in the next section.  We found in \S\ref{kineticcorrections}\ 
that the radiative corrections to the effective action can easily be negligible.  
To see this in the case of (\ref{veff}), we note that $N_3\sim 1$ since $\mu_a/(2\pi g f) \sim (\omega/\mu_a)(\mu_a^2/(g\dot\phi)) \ll 1$ will hold in the regime of frequencies $\omega/H<1$ that we prescribe below.  Given that, $v_{eff}\ll 1$ follows from the fact that $g\dot\phi/\pi\mu_a^2<1$.
The contributions of other diagrams are given by (\ref{ratioa}) which is manifestly subdominant, and (\ref{betaa}).  As noted below (\ref{betaa}), the relevant ratio is less than
\be\label{betaaagain}
e^{\frac{\pi\mu_a^2}{2 g\dot\phi}}\frac{H}{\mu_a}\sim e^{\frac{\pi\mu_a^2}{2 g\dot\phi}} \frac{H}{\dot\phi^{1/2}}\frac{\dot\phi^{1/2}}{\mu_a}\frac{1}{g^{1/2}}
\ee 
where we conservatively put in the minimal $\chi$ mass in the denominator of the second factor on the LHS.  To check if this may be small, we can proceed as follows.  We first impose that the particle production contribution to the power spectrum is in a viable and detectable regime, say setting
\be\label{PPaset}
g^2\frac{\bar n_\chi}{H^3}\frac{\omega}{H}\sim \frac{\omega}{H} g^{7/2} 58^3 e^{-\pi\mu_a^2/g\dot\phi}\equiv 10^{-2}
\ee 
where we have substituted $\dot\phi^{1/2}\sim 58 H$ from the normalization of the power spectrum.  
Having done so, we then evaluate the ratio (\ref{betaaagain}) as a function of $g$.  Implementing this, we find a regime of couplings  where the ratio is $\le 1$, so the novel shape we have calculated is a dominant effect.

Let us next focus on case (b), where the additional structure of the oscillating mass led to a more subtle analysis.  It is useful to collect estimates for the sizes of several quantities we have computed above. 
First, let us denote the exponent suppressing the particle production by 
\be\label{Zdef}
Z=\frac{\pi\mu_b^2}{g\dot\phi}>1
\ee
where we will explain the required inequality shortly.  
We can reduce the parameter space to the three directions $Z, g, \omega/H$  using the relations
\be\label{simpparam}
\mu^2=\mu_b^2+2 g^2 f^2, ~~~ f=\frac{\dot\phi}{\omega}, ~~~ \dot\phi = \pi \frac{\mu_b^2}{g Z}, ~~~ \frac{\dot\phi}{H^2}\simeq 58^2
\ee
(the last one following from the normalization of the power spectrum).  

First, the relative modification of the power spectrum is
\bea\label{SN2}
\frac{\delta\expect{\zeta^2}}{\expect{\zeta^2}} &\sim&\pi\frac{\bar n_\chi}{H^3}g^4\frac{f^2}{\mu^2}\frac{H}{\omega}
\lesssim 10^{-2} \nonumber \\
\eea
where in the last step we enforce that this is within observational bounds.
Next we write the scaling of the $N=3$ point correlator at order $|\beta|^2$ generated by 3-point vertices ${\cal L}_3$ as derived in \S\ref{sec:sourced} (dropping the $(2\pi)^3\delta(\sum \k)$)
\bea\label{leadingbetasq}
\delta\phi^3_{|\beta|^2}k^6&\sim &\pi ^{3/2} \left(\frac{H}{\omega}\right)^{1/2} c_b^3 \bar n_\chi\,.
 \nonumber 
\eea
 The ratio of the contributions from higher point vertices to this contribution is (a power of) the quantity (\ref{ratio}), which must be less than 1 in order for our new shape to be relevant.
Next, we record the leading contribution at order $\beta$, which is  dominated by the contribution of a single insertion of the ${\cal L}_5$ vertex (\ref{shapesaddle}):        
\bea\label{threebeta}
\delta\phi^3_{\beta}k^6&\sim& \frac{\pi^{1/2}}{2^3} H^3 \left(\frac{\bar n_\chi}{H^3}\right) e^{Z/2}\left[\frac{2\mu\pm \omega}{H}\right]^{5/2} \left(\frac{g^2 f^2}{2\mu H}\right) \left(\frac{H^6}{(2\mu\pm\omega)^3f^3}\right)\frac{1}{\left(1+ \frac{g\dot\phi}{\pi \mu_b H}\right)^{3/2}}\frac{\omega}{H} \nonumber \\
\eea
where we have conservatively taken $k\eta_n\sim k \eta_*$, and include a factor of $\omega/H$ for the number of terms contributing at the same level to the sum in (\ref{shapesaddle}).  
The ratio of this to (\ref{leadingbetasq}) must satisfy
\be\label{betabetasqratio}
\frac{{\cal O}(\beta)}{{\cal O}(|\beta|^2)}
<1
\ee
again from the requirement that the novel shape we have computed contributes a leading effect.

 In order to replace the post-production value of $\omega_\chi(\phi)$ by $m_\chi(\phi)$ in our calculations above, neglecting the momentum squared of the produced particles $\sim g\dot\phi/\pi$ relative to $m_\chi^2$,   we should impose $\mu_b^2>g\dot\phi/\pi$, or equivalently $Z>1$.   Given this, we automatically satisfy the condition $\mu_b^2>0$ from (\ref{gfmuineq}-\ref{minmassb}).  
We also impose
\be\label{gfmuconds}
\frac{g f}{\mu_b}>\frac{\sqrt{2}}{\pi}
\ee
(for convenience) so that the particle production timescale be less than half a period of our shift symmetry along the $\phi$ direction.  

We should also comment on the ratio $2 g^2 f^2/\mu^2$ appearing in the mass formula (\ref{bmasssquaredgf}) and (\ref{massexpand}).  
As discussed above, when this ratio is small, the source is a pure sinusoidal function.  However, for the lower end of our frequency range of interest, we find that once we impose that (\ref{SN2}) is in the viable and detectable range, this ratio has a minimal value not hierarchically smaller than 1 (e.g. $\sim .9$ for $\omega/H \sim 10$, and even larger for lower frequencies).  However, as discussed above, the leading Fourier coefficient in (\ref{massexpand}) is similar to that for the pure $\sin(\omega t)$.
Again, one might expect higher harmonics to be suppressed in their contribution, as they oscillate more rapidly and lead to greater suppression from the $1/(k\eta_n)^3$ factor, but it may be worthwhile to include a small number of additional terms in the Fourier series. 

We also have the condition (\ref{smallrhochi}), which reduces to
\be\label{backreaction}
\frac{\rho_\chi'}{V'}\sim \frac{g\bar n_\chi}{3 H\dot\phi}
\ll 1
\ee
which is easily subdominant, as well as (\ref{smallBR2}).   We must also satisfy the condition (\ref{veff}), which reduces to
\be\label{vsq}
\frac{g^2N_3}{2 Z^2} \ll 1
\ee

By reducing the parameter space as in (\ref{simpparam}) and evaluating these quantities numerically, we find that these conditions can all be satisfied in an interesting window
of parameters, ranging over all $\omega/H$ of interest, i.e. ${\cal O}(0.1)<\omega/H<{\cal O}(100)$, with marginal solutions to some of the conditions at higher frequencies. 



In the space of solutions to these conditions, the ratio $(S/N)_3/(S/N)_2$ can easily be order 1 (or larger in some regimes).   In the next subsection, we will discuss the parametric reason for the greater strength of non-Gaussianity derived from our particle production effect as compared to that in \cite{resonantNG, resonantEFT}. 
It would be interesting to understand if there is a more subtle breakdown of the theory that limits the strength of non-Gaussianities.   



%


\subsection{Amplitude of bispectrum versus power spectrum}\label{bispectrumamplitude}

At this point it is interesting to compare the strength of our signal with those coming purely from a sinusoidal term in the inflaton potential.  The latter leads to oscillatory contributions to the power spectrum and resonant non-Gaussianity \cite{resonantNG}\cite{monodromy}\cite{resonantEFT}.  In an effective field theory treatment, the ratio of signal to noise in the bispectrum to that in the power spectrum scales like $\omega^2/M_*^2$, with $M_*$ the strong coupling scale of the EFT \cite{resonantEFT}.  If one started with a purely non-Gaussian oscillatory interaction at the level of the effective theory of the perturbations, it generates a generically larger signal in the power spectrum via a loop correction \cite{resonantEFT}.   Starting from the canonical example of an oscillatory term of the form $\Lambda^4\cos(\frac{\phi}{f})$ in a slow-roll potential, we obtain the simple estimate
\be\label{three2tworesonant}
\left.\frac{(S/N)_3}{(S/N)_2}\right|_{resonant}\sim \frac{{\cal L}_3}{{\cal L}_2}\sim \frac{\delta\phi}{f}\sim \frac{\omega}{f}\sim\frac{\omega^2}{\dot\phi}\sim \left(\frac{\omega}{H}\right)^2\frac{H^2}{\dot\phi}\sim 10^{-5}\left(\frac{\omega}{H}\right)^2\,,
\ee      
where we used the fact that the modes are created at the time when their energy is $\omega$, rather than $H$ as for vacuum fluctuations.  This ratio is less than 1 when the low energy theory is weakly coupled.   For generic $\omega/H$ below this, the signal/noise in the power spectrum is greater than that in the bispectrum.   As noted in \cite{PlanckInflation},
a  coincidence of such frequencies in the two, at a similar signal/noise, could not be explained by this theory. 

In contrast, in the present analysis, as just explained (\ref{SNR}), we obtain (for case (b))
\be\label{three2two}
\left.\frac{(S/N)_3}{(S/N)_2}\right|_{(b)}\sim c_b\sqrt{\frac{\omega}{H}}\sim \left(\frac{g^4 f^2\omega}{\mu^2 H}\right)^{1/2}\,,
\ee
with $c_b$ defined above in (\ref{cb}).  

This ratio is very interesting as it determines the relevance of non-Gaussianity searches and joint analysis with searches for related structure in the power spectrum.    Since it grows with the coupling, we might expect it to be constrained by radiative stability.    It is straightforward to check, however, that the conditions we developed in section 2 do not require the ratio (\ref{three2two}) to be parametrically $<1$.  For example, the conditions \eqref{gfmuineq} and \eqref{g} just require this ratio to be $<\sqrt{\omega/H}$.  Plugging in the numbers and imposing all our control conditions, and the conditions for dominance of the new shapes,  we find a viable range with (\ref{three2two}) of order 1 or somewhat greater (as well as other regimes where it is $<1$).
  
One intriguing result of our analysis is that the $S/N$ in the tree-level $N$-spectra increases with $N$ for a range of model parameters.  This result is perhaps surprising from the point of view that intuitively, a weakly coupled theory should not develop large non-linearities. However, it is important to note that this growth does not persist for arbitrarily large $N$, since the signal to noise
\be\label{SNN}
N^{-1/2}_{\rm modes}(S/N)_N \sim \frac{\langle \zeta^N\rangle}{\sqrt{\langle\zeta^N \zeta^N\rangle}}
\ee  
saturates at order 1 when the denominator is dominated by the non-Gaussian $2N$-point function.  (Before that, the denominator is approximated by $\langle\zeta^2\rangle^{N/2}$ as in (\ref{SNR}).)

It will be very interesting to develop further theoretical and data analysis techniques for the regime with increasing $S/N$ \cite{Moritz}, perhaps using the Poisson statistics and real space distribution of the underlying production events \cite{MSSZ}.

\section{Templates for the power spectrum and bispectrum and parameter ranges}\label{templates}

Having calculated the correlators above and their regime of validity, we are finally in a position to lay out templates for data analysis, assess their overlap with previously searched shapes, and specify parameter ranges.
We will prescribe this in the general case as well as our specific examples (a) and (b).  In these specific cases, we have developed the theoretical consistency conditions in detail in this work, and can prescribe parameter ranges by combining these with the overlap of the new shapes with previously constrained templates.  In case (a), the latter is highly constraining:  it overlaps strongly with equilateral non-Gaussianity except at the lowest frequencies.  In case (b), we find that both the theory and the (weak) overlaps with other templates motivate a search over a broad range of frequencies.    

\subsection{General case}
In terms of the definitions~\eqref{Gsimp},~\eqref{hdef}
\bea\label{green}
\hat g(y) &=& \sin(y)-y\cos(y)\,,  \nonumber \\
\hat h_a(k\eta_n) &=& \int_{\eta_n}^0 \frac{d\eta'}{\eta'}\hat g(k\eta')\frac{\delta m_\chi}{\delta\phi}|_{\phi=\phi_0(t)}\,, \nonumber \\
\eea
and the conformal times of the production events
\be\label{etan}
\eta_n=-\frac{1}{H} e^{\frac{2\pi H}{\omega}(n+\frac{\gamma}{2\pi})}\,,
\ee
where we include a phase parameter $\gamma\in (0, 2\pi)$, the power spectrum can be approximated by the shape
\bea\label{powerspectrum}
\langle\zeta_k\zeta_k\rangle' 
&\simeq&\langle\zeta\zeta\rangle' |_{vac}+ \frac{A_2}{k^3}\sum_{n=\hat n_{min}}^{\hat n_{max}}\frac{\hat h(k \eta_n)^2}{-k^3\eta_n^3} \,.
\eea
The prime denotes dropping the $(2\pi)^3\delta(\k+\k')$, and the first term on the RHS is the usual vacuum contribution.  The bispectrum can be approximated by the shape
\bea\label{bispectrum}
\langle \zeta_{k_1}\zeta_{k_2}\zeta_{k_3}\rangle'
&\simeq& \frac{A_3}{k_1^2k_2^2 k_3^2} \sum_{n=\hat n_{min}}^{\hat n_{max}} \prod_{i=1}^3 \frac{\hat h(k_i\eta_n)}{-\eta_n k_i}\,,
\eea
where again we drop $(2\pi)^3\delta(\k_1+\k_2+\k_3)$. 

As we explained, the dominant contributions for a mode arises when it crosses the horizon so that only a finite number of terms contribute. We will work with fixed endpoints of the sum. The $k$-independent integers $\hat n_{min}$ and $\hat n_{max}$ should satisfy
\bea\label{nhatlimits}
\hat n_{min} &\ll &    \frac{\omega}{2\pi H}\log\frac{H}{k_{max}}\,,\\
\hat n_{max}&\gtrsim &  \frac{\omega}{2\pi H}\log\frac{p_{max}}{k_{min}}\,,\nonumber\\ 
\eea
with
\be\label{pmaxsize}
\frac{p_{max}}{H}\sim \zeta^{-1/2}\sim 10^{5/2}\,.
\ee
They are chosen so that they cover the needed range
for the whole range of $k$ considered in the analysis.  In the shape functions given above, the summand automatically shuts off the sum at the required, $k$-dependent values at both ends.   For each $k$, a total number of terms at most of order $\Delta n\sim \frac{\omega}{2\pi H}\log(\frac{p_{max}}{H})$ contributes to the sum, reflecting the overall scale invariance of the setup.  But the $1/\eta_n^3$ suppresses the early values of $\eta_n$, so in fact the dominant contributions come from the smallest values of $k_i\eta_n$ such that the integral in (\ref{green}) gets its dominant contribution (including its resonance in cases with oscillation in the $\delta_\phi m_\chi (\phi_0(t))$ factor).

\subsubsection{Case (a)}

In case (a), $\frac{\delta m_\chi}{\delta\phi}$ is a step function.  Hence the source integrated against the Green's function  (\ref{green}) becomes
\be\label{hhata}
\hat h_a(k\eta_n) = \int_{\eta_n}^0 \frac{d\eta'}{\eta'}\hat g(k\eta')
\ee
By evaluating the appropriate overlap \cite{shapes}, we find that unless $\omega/H\lesssim\mathcal{O}(1)$, shape (a) is essentially equilateral with tiny oscillations. We show the the shape along $k_1=k_2=k_3=k$ in figure~\ref{plots2a}.
The parameters and ranges for this case (a) are therefore:
\bea\label{parametersa}
{\rm phase:}~~~~~\;\hskip .5mm 0 &\le& \gamma < 2\pi\\
{\rm input ~ frequencies:} ~~~~ \frac{1}{10} &<& \frac{\omega}{2\pi H} < 1  \\
{\rm UV ~ scale:}~\frac{p_{max}}{H} &\simeq& \zeta^{1/2} 
 \eea
and we should again note that a multifrequency analysis as in \cite{Planck15NG}\ would also be appropriate.

\subsubsection{Case (b)}

In this case we have sinusoidal oscillations in $\frac{\delta m_\chi}{\delta\phi}$ (\ref{massexpand}), so  
\be\label{hhatb}
\hat h_b(k\eta_n) = \int_{\eta_n}^0 \frac{d\eta'}{\eta'}\sin\left(\frac{\omega}{H}\log(\eta'/\eta_n)\right)\hat g(k\eta')+{\rm higher~ Fourier~ modes}
\ee
The integral can be performed exactly or in a saddle point approximation (\ref{hbint}) which is a good approximation at sufficiently high frequencies.

As discussed below (\ref{massexpand}), one or more of the higher Fourier modes may be interesting to include, although they are somewhat suppressed within the resonant integral.   This entails a set of frequencies with a specific relation between them:  $\omega_N=N\omega$ where $N=1, 2, \dots$.  Such modes would arise in a more general study of periodic mass functions \cite{Moritz}.    

In this case, the theoretical parameter space includes a viable window satisfying the consistency conditions delineated above, for the full range of $\omega/H$ that can be described in effective field theory.  This includes a regime with competitive signal to noise in the bispectrum as compared to the power spectrum. Moreover, the overlap between the shape generated by (\ref{hhatb}) and the equilateral shape is small (percent level).  Hence the parameter window we propose for this search is
\bea\label{parametersb}
{\rm phase:}~~~~~\; \hskip .5mm 0 &\le& \gamma < 2\pi\\
{\rm input ~ frequencies:} ~~~~ \frac{1}{10} &<& \frac{\omega}{H} < 200  \\
{\rm UV ~ scale:}~\frac{p_{max}}{H} &\simeq& \zeta^{1/2} 
 \eea
In this case as well, a multifrequency analysis would be well motivated.  A specific example of that would be to include more Fourier modes in (\ref{hhatb}), which involves frequencies that are multiples of the lowest frequency.  More generally, multifield versions of the dynamics we studied in this paper could generate a similar pattern with additional frequencies.


\subsection{Shapes and overlaps}

Let us also briefly illustrate the shapes and their comparison to previous templates.  For the latter, we implement the prescription the overlap of signal to noise in different templates developed in \cite{shapes}, applied to our situation which is not scale invariant  (since there are oscillations as a function of the overall $k$ scale).\footnote{We thank M. Munchmeyer for sharing his independent analysis of this, as well as an additional check that the shape is also orthogonal to the resonant shape.}

\subsubsection{Plots for case (a)}

\begin{figure}[htbp]
\begin{center}
\includegraphics[width=6.7in]{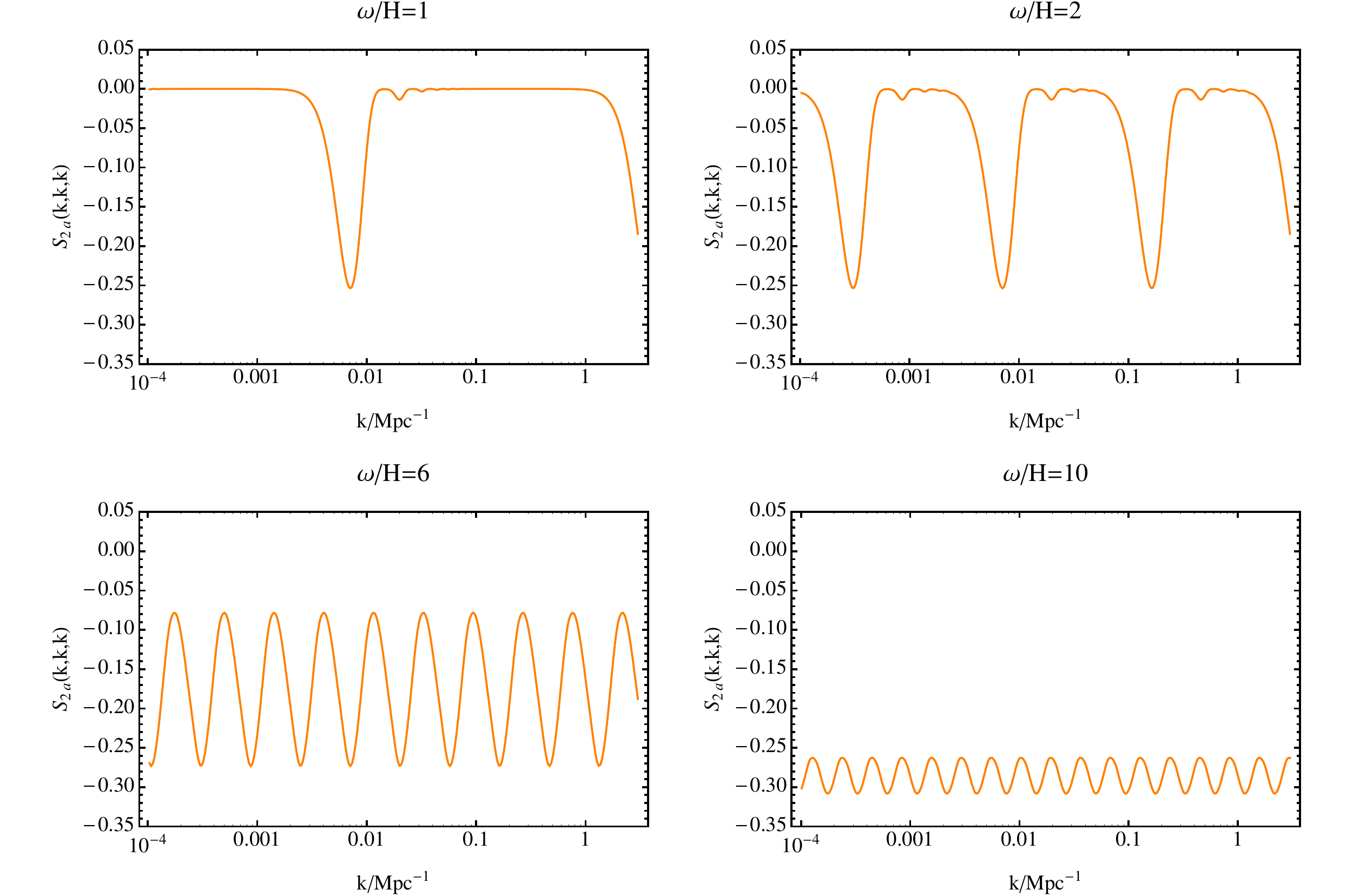}
\end{center}
\caption{Bispectrum shape (a) plotted along the equilateral axis for a range of frequencies. As the frequency increases a plateau develops and the amplitude of the oscillations decreases. \label{plots2a} }
\end{figure}

In figure~\ref{plots2a}, we show the shape $S(k_1,k_2,k_3)=k_1^2k_2^2k_3^2 B(k_1,k_2,k_3)$ for equilateral triangles and a range of frequencies $\omega/H$.

\begin{figure}[htbp]
\begin{center}
\includegraphics[width=4.5in]{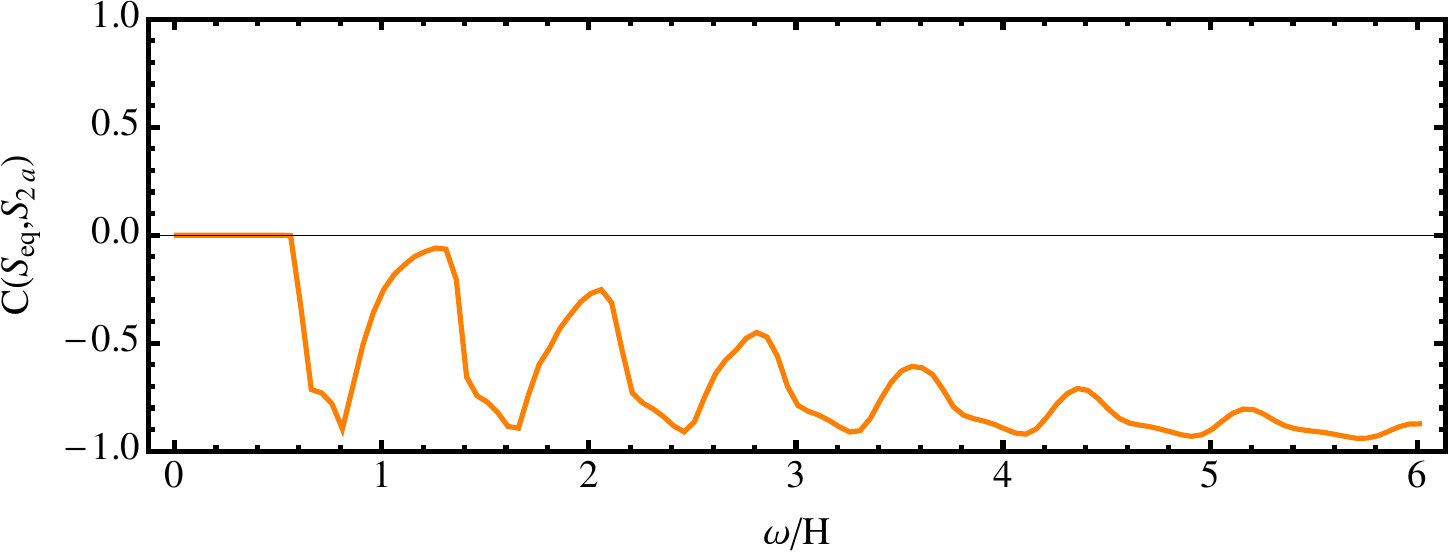}
\end{center}
\caption{Overlap of shape (a) for $\gamma=0$ with the equilateral template using the prescription developed in \cite{shapes}. As the frequency increases the shape approaches the equilateral shape.\label{cos2a} }
\end{figure}

As the frequency $\omega/H$ increases the shape in this direction consists of oscillations on top of a plateau. 
%
%
If we multiply the curve in these figures by~$k^2$, we obtain a shape of the signal that takes into account of the signal-to-noise ratio.\footnote{This is different than for a $3d$ survey, where the signal to noise grows at high k's as $k^3$.}  As one might expect from figure~\ref{plots2a}, overlap calculations confirm that this shape becomes similar to equilateral at $\omega/H>{\cal O}(1)$. This is shown in figure~\ref{cos2a}


\subsubsection{Plots for case (b)}
\begin{figure}[htbp]
\begin{center}
\includegraphics[width=6.7in]{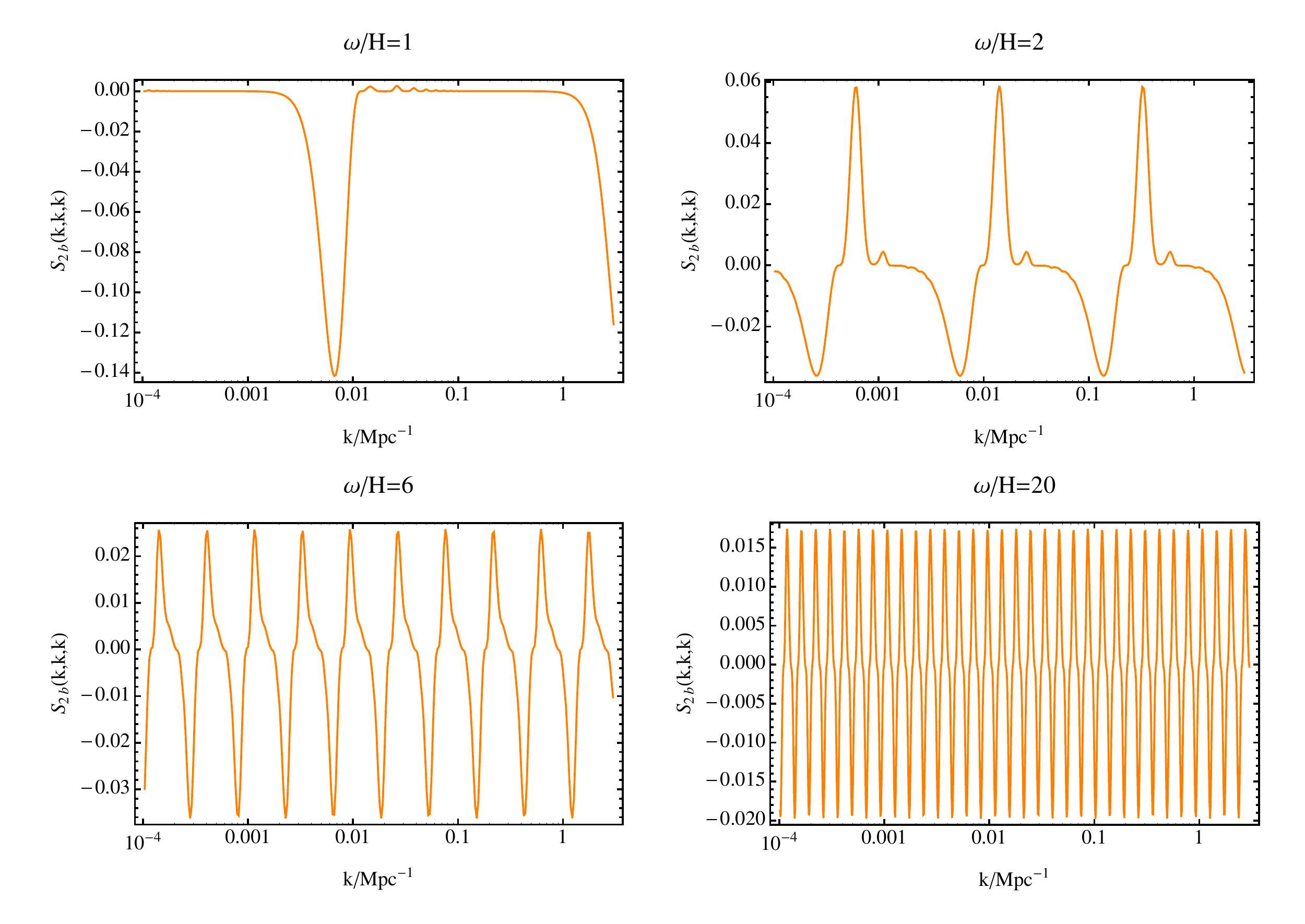}
\end{center}
\caption{Shape for case (b) plotted along the equilateral axis for a range of frequencies.  \label{Sofk2b} }
\end{figure}

\begin{figure}[htbp]
\begin{center}
\includegraphics[width=4.5in]{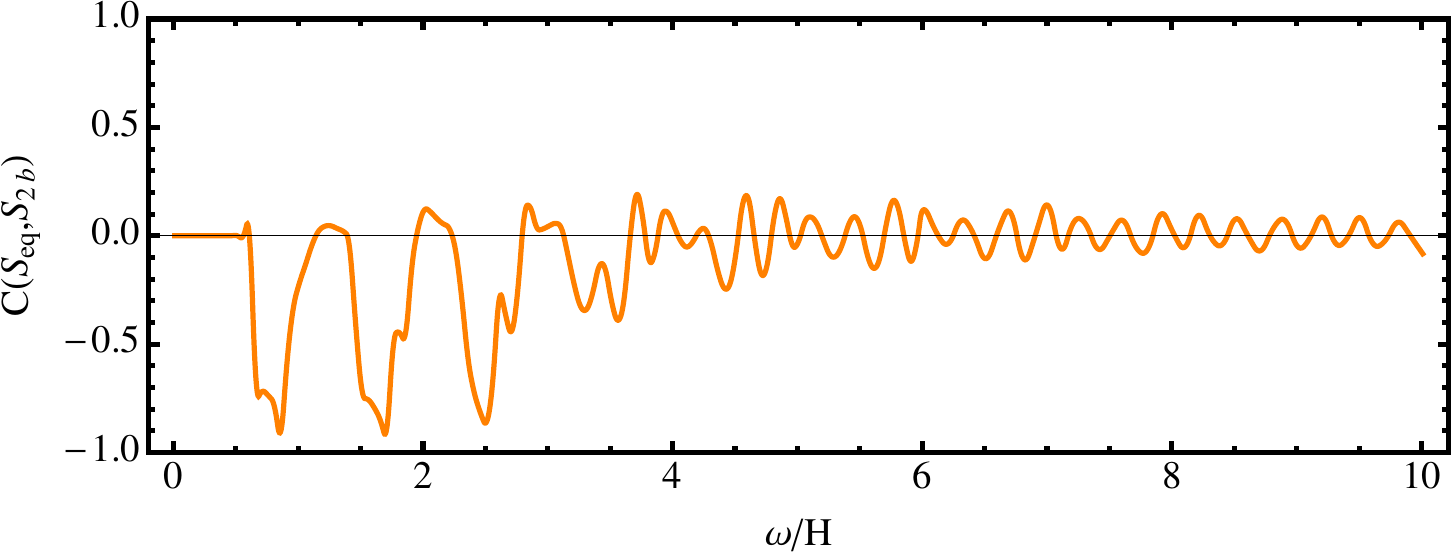}
\end{center}
\caption{Overlap of shape (b) with the equilateral template using the prescription developed in \cite{shapes} \label{plot2b} }
\end{figure}

Figure~\ref{Sofk2b} shows the shape of the bispectrum for case (b) over a range of frequencies for equilateral triangles. 
In this case, the shape oscillates around zero. As a consequence, the overlap with the equilateral shape is low for a wide frequency range. This is shown in figure \ref{plot2b}.  

\section{General lessons}\label{general}

Having developed this mechanism in detail, let us put it in context of the broader study of primordial non-Gaussianity and power spectrum features.
A variety of examples demonstrated that the dynamic range of inflation and its signatures extends well beyond the single-field slow roll case (see for example \cite{SalopekBond, modulated, DBI, trapped, Chen}).    Further progress was made by systematically characterizing the observables from an effective field theory perspective \cite{EFT}.     In this section, we will explore a potentially more systematic approach to the subject.  


\subsection{Dynamics}  Intuitively, in single-field slow roll inflation, the flatness of the potential translates into small interactions and hence Gaussian perturbations.   Non-Gaussianity can be relatively large in many circumstances that are also straightforward to understand.    Dynamical mechanisms for non-Gaussianity can be broadly classified as follows.\footnote{Combinations of these are of course also possible and arguably more generic; see \cite{disorder}\ for interesting proposals for testing a random set of effects using ideas from disorder and localization theory.}

\smallskip

\noindent (1)  Interactions slow the evolution of the inflaton down its potential, which can be steep.  These interactions can naturally lead to detectably large non-Gaussian perturbations.  

\smallskip

\noindent (2) Even in slow-roll inflation, multiple light fields can fluctuate significantly during inflation, with those transverse to the inflationary trajectory having stronger self-interactions since they are unconstrained by the flatness of the potential along the inflaton direction.  

\smallskip

\noindent (3) Various features in the potential and kinetic terms may lead to modulations of the power spectrum and non-Gaussian correlators.  If the $S/N$ in the non-Gaussianity is sufficiently competitive with that in the power spectrum, this can lead to distinctive non-Gaussian signatures.  

\smallskip

In this work we have developed another mechanism:

\noindent (4)  Non-adiabatic effects arising from the coupling of the inflaton to even very heavy fields can generate detectable non-Gaussianity (including in slow roll inflation with a single lighter-than-Hubble field).

\subsection{Effective field theory treatment}

Let us make some further comments on the effective field theory (EFT) approach to the systematic study of non-Gaussianity.  This is important for two almost contrary reasons:  (i) it determines how symmetry constrains observables and (ii) it clarifies what is left undetermined by EFT considerations.   The EFT of single and multifield inflationary perturbations ~\cite{EFT,Senatore:2010wk,Noumi:2012vr} was constructed by noticing that the epoch of inflation can be thought of as a period where time-translations are spontaneously broken and there is therefore a Goldstone boson, {\it i.e.} a light degree of freedom that non-linearly realizes this symmetry, usually denoted $\pi$. The Lagrangian for the fluctuations can therefore be constructed without knowledge of the symmetry breaking mechanism. Additional fluctuating degrees of freedom, if relevant, can be consistently coupled to this Goldstone boson~\cite{Senatore:2010wk,Noumi:2012vr}. In the case that only the Goldstone boson is relevant, and in the so-called decoupling limit where metric fluctuations can be neglected, the EFT Lagrangian takes the schematic form~\cite{EFT}
\bea\label{EFTLagrangian}
&&S=\int d^4x\,\sqrt{-g}\; \left[\mpl^2\dot H(t+\pi)\left(\dot\pi^2-\frac{(\d_i\pi)^2}{a^2}\right)+\right.\\ \nonumber
&&\left.\vphantom{\frac{(\d_i\pi)^2}{a^2}}\qquad+M^4_1(t+\pi)\left(\dot\pi^2+\dot\pi^3+\ldots\right)+M^4_2(t+\pi)\left(\dot\pi^3+\ldots\right)+M^4_3(t+\pi)\left(\dot\pi^4+\ldots\right)\right]\ .
\eea
Here we have neglected operators that are higher order in derivatives and the number of the fluctuations, which are irrelevant for the discussion\footnote{See for example~\cite{Behbahani:2014upa} for a detailed list of operators in the case of single field, and~\cite{Senatore:2010wk,Noumi:2012vr} for multifield}. $M_{1,2,3,\ldots}$ are free functions of time that are not constrained by the symmetry of the problem, in addition to $H$ and $\dot H$, which have to obey only the mild constraint that the spacetime must be inflationary. This implies that there are infinite series of operators in $\pi$ that are unspecified.

{\bf Symmetries:}  One can classify the results in terms of the level of symmetry, which when present can constrain the behavior of these functions. A natural case to consider is a continuous shift symmetry for $\pi$, which requires the unspecified functions of time to simply be constant~\cite{EFT}. Another possibility is a discrete shift symmetry, {\it e.g.} a sinusoidal time-dependence~\cite{resonantEFT}, as it already appeared in axion models~\cite{monodromy}. In these cases, there is a limited number of operators that contribute at a given order in the fluctuations. This has led to a classifications of the possible shapes generated in such symmetric classes of single field inflation, which has been looked for at various stages in the CMB data along with certain multifield shapes (see for example~\cite{Planck15NG,Behbahani:2014upa,Smith:2015uia}).  In general, for multifield inflationary perturbations the possibilities proliferate.    

{\bf General Shapes:} In general, the minimal set of symmetries in single-clock inflation
allow for the unspecified functions of time that are present in the above Lagrangian (\ref{EFTLagrangian}) to be indeed unconstrained. As mentioned, this leaves an infinite set of  operators unspecified that, naively, makes a systematic exploration of the signatures of inflation, even in the purely single field case, hopeless. However, the situation is not as bad as it appears for two reasons. First, primordial fluctuations are small and quite Gaussian, which means that we can restrict to operators with a limited number of fluctuations. This means that we can restrict our analysis to a {\it finite} set of functions, of which we care of the value of a {\it finite} number of time derivatives (for example, $M_{1}^4(t),M_{1}^4{}'(t),M_{1}^4{}''(t),\ldots$)~\footnote{\label{footnote:exception}A potential exception to this statement is the following. One could imagine that there is a set of operators of the Goldstone Lagrangian for which the perturbative expansion in the number of fluctuations is not applicable. If these operators are all small, then the theory would be still approximately Gaussian, but if at the same time there is no relative hierarchy between the operators, one cannot truncate the expansion. Something of this sort happens in the EFT of a particle that is obtained after integrating out a weakly coupled heavy particle: as we push the EFT up to the energy of the particle having been integrated out, the derivative expansion of the EFT breaks down (schematically: $g^2\frac{1}{\Box+m^2}\sim \frac{g^2}{m^2}\left(1-\frac{\Box}{m^2}+\left(\frac{\Box}{m^2}\right)^2+\cdots \right)$). This can be detected in the EFT by 
noticing that the theory is non-perturbative and higher derivative operators are important, signaling the presence of additional degrees of freedom. In the EFT of single-clock inflation, the same phenomenon can appear in two ways: either the derivative expansion breaks down (in which case the theory cannot be regarded as single field), or the expansion of the functions $M_{i}^4(t+\pi)$ breaks down. In this second case, there is not much we can do at the level of the EFT: the operator is strongly coupled and we are unaware of techniques of how to compute it (it is unclear if performing the calculation in unitary gauge would help; we are also unsure if this can ever happen by preserving in the Lagrangian only one degree of freedom. If instead there were to emerge additional degrees of freedom, we could use the techniques developed in~\cite{LopezNacir:2011kk,Green:2013rd}).}. Second, the duration of inflation is {\it finite}, and experiments have a {\it finite} volume coverage. This implies that each function can be expressed as a {\it finite} superposition  of plane waves. For example:
\be
M_I^4(t)=\int d \omega\; e^{i\,\omega\, t}\, \tilde M^4_I(\omega)\simeq \sum_{J=-J_{\rm max}}^{J_{\rm max}}\,  e^{i\,\Delta\omega\, J\, t}\, \tilde M^4_{I,J}
\ee 
The smallest step in frequency is determined by the duration of inflation $\sim N_e H^{-1}$, with $H\gtrsim 10^{-12}$GeV and $N_e\sim 60$ being the number of $e$-foldings.  That is, $\Delta\omega\sim H/N_e$. The largest frequency is limited by the experiment volume coverage $\omega_{\rm max}\sim\Delta \omega\, J_{\rm max}\sim H L k_{max}$, where $H$ is the Hubble rate during inflation, and $L$ is the length (shorter or equal to our current Hubble length) extension of the experiment~\footnote{There are additional factors that might limit the actual $\omega_{\rm max}$ of an experiment. For example, in the case of the CMB, the finite width of the visibility function is expected to induce a wash out of the high frequency oscillations.}.  This arises as follows.

Given an experiment of size $L$, the minimum change in momentum $k/a$ is $1/L$.
For our purposes, we need that two modes differing by $\Delta k/a=1/L$ exit the horizon with a difference in time small
enough that the oscillatory piece we are interested in capturing did not complete an oscillation.
Since in a resonant situation the modes are created at $\eta k\sim\omega/H$, we have $\eta_1/\eta_2\simeq 1+\Delta k/k$ for the times $\eta_1$ and $\eta_2$
at which the two modes are created.
This implies that the change in phase in this time period is
\be\label{phasechange} 
\frac{ \Delta \omega j }{H} \log[\eta_1/\eta_2]\sim \Delta \omega j /H  \frac{\Delta k/a}{k/a} <1, 
 \ee
for all $j$'s. This implies $\Delta \omega j_{max}< H L \frac{k}{a} < HL k_{max}/{a}$.  


So even though general functions of time are allowed in the EFT Lagrangian even at the single-field level, we find that apart for a marginal exception described in footnote~\ref{footnote:exception}, only a finite, though potentially large, set of operators needs to be included to fully describe the space of possibilities.  However, as we pointed out in detail in sec.~\ref{EFT}, relying on a single field description amounts to assuming that there are no particles with mass comparable to the highest frequency at which the EFT coefficients have support. If there were to be additional fields at these scales, they would not be describable with just a single degree of freedom, and the possibilities would proliferate (see~\cite{LopezNacir:2011kk} for some first steps to classify these effects as dissipative effects in a single field Lagrangian). 

{\bf Data analysis:} What is the way to analyze this space of possible signals, even at the purely single-field level? Since non-Gaussianities are observationally bound to be small, it is possible to consider the signal as the superposition of the signal induced by each operator taken with the others set to zero. Since each operator has a sinusoidal time dependence, the signal is the sum, with arbitrary coefficients and with the frequencies as described above, of the effects that were for example studied in the case of resonant non-Gaussianities~\cite{resonantNG,resonantEFT}. The latest Planck analysis~\cite{Planck15NG} has analyzed the resonant shapes, but only for the three point function and for a few frequencies. At least naively, an analyses where templates of multiple frequencies are used should be doable and would offer a more general coverage of the signals captured by the EFT of single field inflation.  

However, the large number of parameters may well dilute away some specific signals, a problem which gets exacerbated with multiple fields, including very heavy ones as studied in this work.  Therefore, insight from UV completions remains important for deriving well-motivated searches for particular signatures.  


{\bf Technical Naturalness:} If we allow for generic functions of time in the coefficients of the EFT Lagrangian, the shift symmetry of $\pi$ gets violently broken. It is expected that large radiative corrections will modify these coefficients, but so far as we do not rely on any specific functional form and instead explore a general form for the functions $M_{i}^4(t)$, technical naturalness does not seem an issue.

{\bf Additional degrees of freedom and dependence on high energy scales:}

As we explained above in \S\ref{EFT}, and worked out in detail in the bulk of this paper, the precision of current data requires including very heavy fields as when their non-adiabatic effects are suppressed by $\sim 1/\sqrt{N_{\rm modes}}$.   In that context, it would be worthwhile to undertake a more systematic study of $m_\chi(t)$, expanding it in Fourier components and imposing consistency criteria~\cite{Moritz}.


\section{Summary and future directions}\label{summary}

We have seen that for a well-defined window of parameters, current CMB data is sensitive to non-adiabatic production of particles with time-dependent mass, even if the minimal mass $\tilde \mu\gtrsim \dot\phi^{1/2}$ is much larger than the scale $\sim H$ of the vacuum fluctuations.  The basic reason for this is that the exponential $\exp(-\pi\tilde\mu^2/g\dot\phi)$ (\ref{expform}) can compete with $1/\sqrt{N_{\rm modes}}\sim 10^{-3}$ for a range of $\tilde\mu^2>g\dot\phi$.  

This general expectation survives a detailed derivation of the $N$-point in-in correlation functions in the quantum field theory describing the coupling of the heavy fields to the inflaton, including quantum interference effects going beyond the classical production scenario outlined in the appendix of \cite{MSSZ}.   
Motivated in part by axion monodromy, we focused on sectors of heavy fields whose production events respect a discrete shift symmetry along the inflaton direction.   
For a range of parameters, the resulting contribution to the scalar perturbations yields novel shapes of non-Gaussianity, with an amplitude that can be competitive with or somewhat greater than the corresponding  contributions to the power spectrum.  This contrasts with resonant non-Gaussianity \cite{resonantNG}.\footnote{But see the second reference in \cite{resonantEFT}\ for an interesting exception.}
(There is another range of parameters for which the effect is still visible at least in the power spectrum and the shape is more similar to the resonant shape, in addition to regimes where it would be too small to observe.)   

There is a range of parameters where the signal/noise in the primordial $N$-spectra grows somewhat with $N$. It would be very interesting to formulate an optimal search strategy for this regime \cite{Moritz}, although the bispectrum analysis can always be done.  To our knowledge, this is the first case of such growth, and deserves further investigation.  

Particularly for high-scale inflation, in which there are few orders of magnitude between $H$ and the Planck scale, such massive fields are expected in many extensions of the Standard model, including grand unified theories and string theory; moreover, the novel regime of shapes and amplitudes we have derived applies for couplings not tuned to be smaller than needed for control of radiative corrections.  However, it should be emphasized that the effect can be easily suppressed by considering sufficiently weak coupling; since it is exponentially suppressed, the regime where it is not visible can arise without substantial tuning of parameters.
Nonetheless, the theoretical genericity of heavy fields coupling to the inflaton combined with the precision of modern cosmological data motivates carrying out a search to determine the empirical constraints on the amplitude of this effect.  
We have provided templates for analysis including parameter ranges determined by theoretical consistency and the level of overlap with existing templates.

In general, we find remarkable the sensitivity of the data to microphysical details, albeit limited by the finite number of modes; this remains worth exploiting to the full extent possible.   One generalization of this work would be to check more explicitly the effects of fermion production \cite{Moritz}, particularly since some level of micoscopic supersymmetry helps control radiative corrections generated by vacuum fluctuations of the heavy fields.  

A very interesting but more difficult generalization would be to string (as opposed to particle) production.  This is also motivated in part by axion monodromy, in which discrete parameters determine whether particle or string production would arise in the sectors described by our case (a).  Theoretically, this has several novel features \cite{backdraft}, so we could not immediately apply our current results to this case.  It would be very interesting to see if the string production process can be analyzed with sufficient theoretical precision to derive specific predictions and templates, and if so, whether these overlap substantially with those we have computed here.

\noindent{\bf Acknowledgements}
We would like to thank Hiranya Peiris and Moritz Munchmeyer for extensive discussions.  We are also grateful to Dick Bond, Francois Bouchet, George Efstathiou, Steve Gratton, Daniel Green, Liam McAllister,  Daan Meerburg, and Ben Wandelt for discussions and correspondence.   We are thankful to the Kavli Institute for Theoretical Physics and to the Aspen Center for Physics for hospitality during parts of this project. R.F. was supported in part by the Alfred P. Sloan Foundation. M.M. was supported by NSF Grants PHY-1314311 and PHY-0855425. The work of E.S.~was supported  in part by the National Science Foundation under grant PHY-0756174 and NSF PHY11-25915 and by the Department of Energy under contract DE-AC03-76SF00515. 

\newpage
\appendix

\section{Radiative corrections}\label{app:radiative}

As discussed in the main text, we must ensure radiative stability of our mechanism and its predictions.  
The details of this depends on additional microphysical specifications, such as the effective cutoff on loop momenta and the level of microscopic supersymmetry.  Without formulating a specific model, in this appendix we will check basic criteria for radiative stability, making sure to capture the leading contributions which descend irreducibly from the time-dependence of the $\chi$ (and possible superpartner) masses, since that is intrinsic to our mechanism.    

We will first review the 1-loop effective action in a minimal extension to a supersymmetric model with (complex) boson $\chi$ and superpartner fermion $\psi$, with $\phi$-dependent mass-squared given by $m_\chi^2 = |m(\phi)|^2$ with $m(\phi)$ the complex mass appearing in the superpotential. We focus on the contributions to supersymmetry breaking from the time dependence, leaving out other contributions to spontaneous supersymmetry breaking that would involve additional sectors.  In this calculation, supersymmetry ensures that all corrections are derivative terms.   We will determine the conditions for radiative stability that descend from this irrreducible contribution to SUSY breaking from the time dependence.
Next we will derive the Coleman-Weinberg potential in case (a) and in the presence of hard SUSY breaking mass splittings,  to show how type (b) couplings can be radiatively generated.

\subsection{1-loop effective action with microscopic ${\cal N}=1$ supersymmetry}

At one loop, we can compute the effective action for $\phi$ in components using heat kernel techniques. The one loop contribution can be written as
\begin{multline}
\Delta S_{eff}=-i \lim_{y\to x} \int d^4 x\\
\left[\frac14 \text{tr}\ln(-\partial_x^2-i(\slashed\partial m_1)+\gamma^5(\slashed\partial m_2)+|m|^2)\delta(x-y)
-\ln(-\partial_x^2+|m|^2)\delta(x-y)\right]\,,
\end{multline}
where the trace runs over spinor indices, $m_1$ and $m_2$ are the real and imaginary parts of $m$, respectively, $\partial_x^2$ should be thought of as an operator whereas the derivative in $\slashed\partial m_i$ only acts on the masses. We can formally write this as
\begin{multline}
\Delta S_{eff}=-i\lim_{y\to x}\int d^4 x \int\limits_{0}^\infty \frac{dt}{t}\\\left[\frac14 \text{tr}e^{-t(\partial^2-i(\slashed\partial m_1)+\gamma^5(\slashed\partial m_2)+|m|^2)}\delta(x-y)-e^{-t(\partial^2+|m|^2)}\delta(x-y)\vphantom{\frac14}\right]\,.
\end{multline}
The mass-squared $|m|^2$ cuts off the integral exponentially at for large proper times $t$, but the integral diverges as $t$ approaches $0$ and has to be regulated. We can use proper time regularization and replace the lower limit by $\epsilon$ or use dimensional regularization. Let us make use of the momentum space representation of the $\delta$-function to rewrite the integrand
\begin{multline}
\Delta S_{eff}=-i\int d^4 x\int \frac{d^d p}{(2\pi)^d} \int\limits_{0}^\infty \frac{dt}{t}\\\left[\frac14 \text{tr}\,e^{-t(-(\partial-i p)^2-i(\slashed\partial m_1)+\gamma^5(\slashed\partial m_2)+|m|^2)}
-\vphantom{\frac14}e^{-t(-(\partial-i p)^2+|m|^2)}\right]\,.
\end{multline}
Here derivatives acting on the right are set to zero but derivatives acting on the masses are kept. The momentum $p$ appears from the action of the derivative on the exponential in the momentum space representation of the delta function, more specifically, as an operator $\partial\exp(-ipx)=\exp(-ipx)(\partial -i p)$. It is convenient to rescale the momenta $p=k/\sqrt{t}$ so that
\begin{multline}
\Delta S_{eff}=-i\int d^4 x\int \frac{d^d k}{(2\pi)^d}e^{-k^2} \int\limits_{0}^\infty \frac{dt}{t^{\frac{d}{2}+1}}\\\left[\frac14 \text{tr}e^{-t(-\partial^2+2i k\cdot\partial/\sqrt{t}-i(\slashed\partial m_1)+\gamma^5(\slashed\partial m_2)+|m|^2)}-\vphantom{\frac14}e^{-t(-\partial^2+2i k\cdot\partial/\sqrt{t}+|m|^2)}\right]\,.
\end{multline}
Expanding up to second order in derivatives of the background field $\phi$ we see that the Coleman-Weinberg contribution (zeroth order in derivatives) cancels because of SUSY and the two derivative contribution is
\be
\Delta S_{eff}=-\int d^4 x \frac{1}{2(4\pi)^{d/2}} \int\limits_{0}^\infty \frac{dt}{t^{\frac{d}{2}-1}}e^{-t |m|^2}\partial_\mu m^*\partial^\mu m\,.
\ee
After expanding $d=4-2\epsilon$ in $\epsilon$, we find
\be
\Delta S_{eff}=\int d^4 x \frac{1}{32 \pi^2}\left(-\frac{1}{\bar\epsilon}+\ln\frac{|m|^2}{\mu^2}\right)\partial_\mu m^*\partial^\mu m\,.
\ee
Higher order corrections are also readily obtained by expanding to higher order in $\d$.
Alternatively, we can derive the two-derivative contribution in the superspace formalism~\cite{Kahler}. If we use a momentum space cut-off $\Lambda$ to regulate divergences instead, and add the tree-level contribution the bosonic part takes the compact form
\be
S_{eff}=\int d^4 x \frac{1}{2} f^2(\phi)(\d\phi)^2 -V(\phi),
\ee 
where 
\be
f^2(\phi) =1 + \frac{1}{16\pi^2}\log\frac{e|m|^2}{\Lambda^2}\d_\phi m \d_\phi m^*.
\ee
Next we make the kinetic term standard by using the canonically normalized field
\be
\phi_c \equiv F(\phi),\qquad F(\phi) = \int f(\phi) d\phi.
\ee
This results in the following leading order correction to the potential 
\be
\delta V =  (\phi - F(\phi)) V',
\ee
which we will demand to have a subleading effect on the slow-roll solutions. Let us focus on some concrete models for $m(\phi)$. In case (a), we can consider $m_1 = \mu_a$, $m_2 = g_a(\phi - 2\pi n f)$. Taking $\log (|m|^2/\Lambda^2) \sim 1$, the condition for radiative stability of the slow-roll solution then becomes
\be\label{ga}
\frac{g_a^2}{16\pi^2}\ll 1.
\ee
One way to realize scenario (b) is to consider $m=\mu_0+{\mu}_1\e^{i\phi/f}$ with $\mu_0$ and ${\mu_1}$ related to $\mu$ and $gf$ according to $\mu^2=\mu_0^2+\mu_1^2$ and $g_b^2f^2=\mu_0\mu_1$. In this case, we need 
\be
\frac{\mu_1^2}{16\pi^2 f^2} \ll 1.
\ee
However, the phenomenologically interesting case is when $\mu_0\sim \mu_1$ in which case we recover \eqref{ga} with $g_a\to g_b$.     Imposing also that the radiative corrections not introduce appreciable resonant corrections to the primordial power spectrum leads to weaker conditions.  

\subsubsection{Higher derivative corrections}\label{kineticcorrections}

So far, We focused on the 2-derivative effective action in this calculation.  Higher derivative corrections will also be generated, but will be further suppressed. For instance, at fourth order in derivative we get
\be\label{dphi4}
\frac{1}{16\pi^2} \frac{(\d m)^4}{m^4} \sim \frac{g^4}{16\pi^2\mu^4} (\d\phi)^4,
\ee
where we used $\d_\mu m(\phi) \sim g \d_\mu \phi$.

At strong coupling, such corrections can generate large non-Gaussianity \cite{DBI}.  In general, since these loop effects are power law rather than exponentially suppressed in the minimal $\chi$ mass $\mu$, we should analyze whether they can dominate over our non-adiabatic effects. The term \eqref{dphi4} appears as the first nontrivial correction in a series of higher dimension operators that are generated.  With extended supersymmetry, it is not renormalized, and also appears as the leading correction in a series of corrections tractable at strong coupling.  That is, in the absence of strong accelerations, and with sufficient microscopic supersymmetry, we can read off the kinetic corrections induced by $\chi$ conveniently from the string-theoretic Born-Infeld action for $N_3$ D3-branes, with the field $\phi$ related to the position $r$ of the brane
\bea\label{SDthree}
S&=&-\frac{1}{g^2 (2\pi\alpha')}\int d^4 x \frac{r^4}{R^4}\sqrt{1-\frac{R^4 \dot r^2}{r^4}}\\
&\simeq& \int d^4 x \left\{\frac{1}{2}\dot\phi^2+ g^4 \frac{N_3\dot\phi^4}{4\pi^2 m_\chi^4}+\dots \right\}\,.
\eea
This action is generated by loop corrections in the low energy quantum field theory on the D3-branes.\footnote{Including all the factors in the D-brane action gives the following identifications:  $\{$Yang-Mills coupling $g_{YM}^2\equiv g^2=2\pi g_s$, $R^4=2 g^2 N_3\alpha'^2$, $m_\chi=r/(2\pi\alpha')=g\phi_c\equiv g\phi$$\}$.  Note that we will call the canonical field $\phi_c=\phi$, although in the DBI literature this was not always the case (off by a factor of the string coupling $g_s$).}   In case (a), we should take the effective $N_3\sim \mu/g(2\pi f)$ if this ratio is $>1$, in order to keep all contributions with a mass of order $\mu$.  In case (b), we have only one sector of $\chi$ fields, so $N_3\sim 1$ in that case.

What we will want to do is impose the requirement we get on $Z\equiv \pi\tilde\mu^2/g\dot\phi$ 
from the power spectrum (\ref{SN2}), and see what it implies for the contribution of the kinetic corrections to the background evolution, perturbations and non-Gaussianity.    

This can all be assessed by noting that the DBI corrections are controlled by 
\be\label{csgamma}
c_s=\frac{1}{\gamma}, ~~~~~ \gamma=\frac{1}{\sqrt{1-\frac{g^4N_3\dot\phi^2}{2\pi^2\mu^4}}}\equiv\frac{1}{\sqrt{1-v^2}}\,,
\ee
with
\be\label{veffA}
v^2=\frac{g^2}{2}N_3\left(\frac{g\dot\phi}{\pi\mu^2} \right)^2 \ll 1\,,
\ee
where the last inequality is the condition for these corrections to be neglected.  
We will find that this is indeed small in our parameter window, so that these kinetic corrections are negligible as a contribution to the evolution and perturbations.  

\subsection{Effective Potential from $\chi$ vacuum fluctuations in case (a)}\label{atob}

In string theory, one has both types of $\chi$ sectors, those of cases (a) and (b) above\footnote{Although as mentioned there, sector (a) sometimes consists of strings rather than low energy fields.}  Let us next discuss the periodic potential term generated by integrating out $\chi$ particles of type (a), including the leading effect of a bose-fermi mass splittings.  This in itself could provide the leading sinusoidal potential term entering into case (b):  for couplings which are VEVS of fields, as in string theory, this periodic term then implies periodically varying masses of type (b).   This interplay between the vacuum and non-adiabatic effects of the $\chi$ sectors deserves more systematic study.    

For now, let us explore the structure of the effective action that we obtain from the vacuum fluctuations of the $\chi_n$ fields (\ref{2amasses}) in sector (a) above to see how these effects may generate the sinusoidal correction in a self-contained way. To analyze this, let us work at a point in $\phi$ and work out the Coleman-Weinberg potential
from each $\chi_n$ sector, and sum them up.   

Let us put in a fermion partner $\psi$ for $\chi$, with mass splittings given by
\be\label{masses}
m_{\chi_n}^2=m_n^2+\Delta m_\chi^2, ~~~m_{\psi_n}^2=m_n^2+\Delta m_\psi^2
\ee
with
\be\label{mn}
m_n^2=\mu_a^2+g^2(\phi-\phi_n)^2 =\mu_a^2+g^2(\phi-2\pi n f)^2
\ee
much greater than the squared mass splittings.   That is, we work in a regime where $\mu$ is much greater than the supersymmetry breaking scale; a stronger periodic contribution would arise otherwise which is a little more complicated to compute.  
From these fields we get a one-loop Coleman-Weinberg potential
\be\label{CW}
\sum_{n=-\infty}^\infty \int^{M_*} \frac{d^4 k_E}{(2\pi)^4}\log\left(\frac{k_E^2+m_n^2+\Delta m_\chi^2}{k_E^2+m_n^2+\Delta m_\psi^2}\right)
\ee
Expanding this in the small ratios $\Delta m^2/(m_n^2+k_E^2)$ gives us a leading contribution\footnote{If we had prescribed extended supersymmetry microscopically, with more bosonic and fermionic partners for $\chi$, the analogous formula would be of order $(\Delta m^2)^2$ or smaller, with more inverse powers of the $\chi$ mass-squared.  That case could be analyzed similarly.}
\be\label{quadratic}
(\Delta m_\chi^2-\Delta m_\psi^2)\int \frac{k_E^3dk_E}{8\pi^2}\sum_{n=-\infty}^\infty \frac{1}{\mu_a^2+k_E^2+g^2(\phi-2\pi nf)^2}
\ee
We can evaluate this sum over $n$, giving us
\be\label{donesum}
(\Delta m_\chi^2-\Delta m_\psi^2)\int \frac{k_E^3dk_E (if)}{16\pi\sqrt{k_E^2+\mu_a^2}}\left(\cot\left(\frac{1}{2 g f}[g\phi-i\sqrt{\mu_a^2+k_E^2}]\right)-\cot\left(\frac{1}{2 g f}[g\phi+i\sqrt{\mu_a^2+k_E^2}]\right)\right)
\ee
For sufficiently large $\mu_a/gf$ we can simplify this using
\be\label{simpexp}
\cot(u+iv)-\cot(u-iv)\sim -2i\left(1+e^{-2v}\cos(2u)\right)+{\cal O}(e^{-4v})
\ee
to get a constant piece (subtracted as part of the cosmological constant tune) plus\footnote{performing the last integral by changing integration variable to $\sqrt{k_E^2+\mu_a^2}$).}
\be\label{expcase}
(\Delta m_\chi^2-\Delta m_\psi^2)\cos\left(\frac{\phi}{ f}\right)\int \frac{k_E^3dk_E (g f)}{8\pi\sqrt{k_E^2+\mu_a^2}}e^{-\sqrt{k_E^2+\mu_a^2}/g f}=(\Delta m_\chi^2-\Delta m_\psi^2)\cos\left(\frac{\phi}{ f}\right)\frac{g f\mu_a}{4\pi}K_1\left(\frac{\mu_a}{g f}\right)
\ee
with $K_1$ a Bessel function which behaves as 
\be\label{K1asymptotic}
K_1\left(\frac{\mu_a}{g f}\right)\sim\sqrt{\frac{g f\pi}{2\mu_a}}e^{-\mu_a/gf}
\ee
at large argument.  The more general integral expression (\ref{donesum}) must be used (along with an appropriate subtraction for the cosmological constant) when $\mu_a/g f$ is not large.  

This result is interesting in that it provides a simple mechanism for generating the sinusoidal term in the potential, with an amplitude that depends on the same parameters that will appear in some of the particle production effects.  Its size and its detailed dependence on the parameters can be compared and contrasted with the case where the leading such term is generated by instanton effects.  In some simple situations, such as string-theoretic axions in a single-scale compactification manifold, the instanton effects scale like $e^{-M_P/f}$ as reviewed in \cite{drift}.  The present calculation is more relevant in a different regime, where the $\chi_n$ sectors run in loops in perturbative Feynman diagrams.       

To summarize, we have generated a cosine term from vacuum loops of the $\chi_n$ sectors of type (a) defined above.  In a string-theoretic context, the parameters in the result ($\Delta m, \mu_a, f$) depend in general on additional fields, some of which may play the role of the $\chi$ particles of type (b) above.  In such a situation, the (a) sector would generate oscillating masses for the (b) sector.

\end{document}